\documentclass[11pt,english]{article}

\usepackage[utf8]{inputenc} %

\usepackage{dsfont}

\usepackage[left=1in,right=1in,top=1in,bottom=1in]{geometry}
\usepackage{authblk}
\usepackage{enumitem}%
	\usepackage{ulem} %
	\usepackage{empheq}
	
	\usepackage{graphicx}%
	\usepackage[percent]{overpic}
	\usepackage{subcaption}
	\usepackage{float}%
	\usepackage{wrapfig} 
	
	\usepackage{xcolor}
	\usepackage{color}
	\usepackage{framed}
	\definecolor{darkgreen}{rgb}{0,0.5,0}
	\definecolor{darkblue}{rgb}{0,0,0.6}
	\definecolor{purple}{rgb}{0.4,.2,0.7}
	
	\usepackage{amsmath}%
		\usepackage{amssymb}
		\usepackage{mathrsfs}
	
	\usepackage{tensor} %
	\usepackage{slashed}%
	\usepackage{mathtools}%
	\usepackage{tikz-cd}
	\usepackage{leftidx}%
	\usepackage{esint}%
	\usepackage{amsfonts,amsthm,bm}%
	
	\usepackage{physics}
	\usepackage{braket}

	\usepackage{prettyref}
	\usepackage[colorlinks=true,citecolor=darkgreen,linkcolor=purple,urlcolor=purple]{hyperref}%
	
	\usepackage[numbers,sort&compress]{natbib}

	\newcommand{\nn}{\nonumber}

	\definecolor{burntorange}{HTML}{F7921D}

	\numberwithin{equation}{section}
	\numberwithin{figure}{section}
	\numberwithin{table}{section}
	
\begin{document}
		
		\title{\textsc{{\LARGE  De Sitter  Observers and\\
				 an Effective Field Theory of Horizon Edge Modes}
			}}

			\author[a]{\normalsize James Bonifacio}
		\affil[a]{ \it \normalsize	 Department of Physics and Astronomy,
			University of Mississippi, University, MS 38677, USA}
			\author[b]{\normalsize Kurt Hinterbichler}
		\affil[b]{ \it \normalsize	 CERCA, Department of Physics,
			Case Western Reserve University, 10900 Euclid Ave, Cleveland, OH 44106, USA}
			\author[c]{\normalsize Y.T.\ Albert Law}
		\affil[c]{ \it \normalsize	 Leinweber Institute for Theoretical Physics at Stanford, 382 Via Pueblo, Stanford, CA 94305, USA}
		\author[d]{\normalsize Varun Lochab}
		\affil[d]{ \it \normalsize	 Center for Theoretical Physics, Columbia University, New York, NY 10027, USA}
		\date{}
		\maketitle
		
		\begin{center}
			\vskip-10mm
			{
				\href{mailto:bonifacio@phy.olemiss.edu}{bonifacio@phy.olemiss.edu},
				\href{mailto:kurt.hinterbichler@case.edu}{kurt.hinterbichler@case.edu}, 
				\href{mailto:ytalaw@stanford.edu}{ytalaw@stanford.edu}, \href{mailto:vv2338@columbia.edu}{vv2338@columbia.edu}
			}
		\end{center}

		\vskip10mm
		
		\thispagestyle{empty}

		\begin{abstract}

			We show that choosing a test observer in $dS_D$, specified by the framed embedding of their horizon $S^{D-2}$, spontaneously breaks the de Sitter and internal sphere isometries to diagonal rotations. We identify gravitational edge modes, described by low diffeomorphism jet fluctuations near the horizon sphere, as the associated Goldstone modes. We construct their membrane-like effective theory, organized in a small-gradient expansion. We find a class of nonlinear actions that at the quadratic order reproduce the edge-mode spectrum previously found in the one-loop gravitational $S^D$ path integral.

		\end{abstract}
		
		\newpage
		
		\tableofcontents

		\section{Introduction}

		A key distinction between a black hole horizon and a de Sitter (dS) horizon  is that the latter is intrinsically tied to an observer. In dS space, different (inertial) observers have access to different static patches. These patches are related by de Sitter isometries, but a fixed patch is preserved by only a subgroup. This prompts the question: how is the full de Sitter symmetry realized in an observer-dependent static-patch description of gravity?
		
	A hint arises from interpreting the Euclidean sphere path integral as a thermal partition function in a \textit{Lorentzian} $dS_D$ static patch \cite{Gibbons:1976ue}. At one loop, the gravitational partition function on $S^{D}$ admits a bulk-edge factorization \cite{Anninos:2020hfj,Law:2020cpj,Law:2025ktz,Law:2026tuk},\footnote{Factorization of the form \eqref{eq:Z1loopbulkedge} extends to more general 1-loop Euclidean black hole path integrals
		\cite{Grewal:2022hlo,Kapec:2024zdj,Law:2025yec,Mukherjee:2025xlt}.} 
		\begin{align}\label{eq:Z1loopbulkedge}
			Z^\text{1-loop}_\text{grav} = Z_\text{bulk} Z_\text{edge} \;. 
		\end{align}
	Here, $Z_\text{bulk}$ is the thermal partition function for an ideal graviton gas in a $dS_{D}$ static patch \cite{Law:2022zdq,Law:2023ohq}. The remaining factor, \(Z_{\mathrm{edge}}\), is expressed in terms of functional determinants on an \(S^{{ D-2}}\) of the dS radius $\ell$, naturally identified with the horizon sphere. Explicitly, \cite{Law:2025ktz,Law:2026tuk} 
	\begin{align}\label{eq:Zedgegravresult}
		Z_{\rm edge }& \propto \det\nolimits'\!\left| -\nabla_0^2-\tfrac{{ D-2}}{\ell^2}   \right|  \times \det\nolimits'_{-1}\!\left| -\nabla_1^2-\tfrac{{ D-3}}{\ell^2} \right|^{\frac12}
		\times \det\nolimits'\!\left(-\nabla_0^2 \right)^{\frac12} \; .
	\end{align}
	Here $-\nabla_0^2$ and $-\nabla_1^2$ denote the scalar and transverse-vector Laplacians on  the round sphere $S^{{ D-2}}$ of radius $\ell$.\footnote{Primes indicate omission of zero modes, while the subscript ``$-1$’’ denotes that the harmonic product is extended to begin at degree $l=-1$.} Expression \eqref{eq:Zedgegravresult} makes manifest the ghost-like contributions: a tachyonic vector $A_i$, two tachyonic scalars $\phi^a$ ($a=0,D$), and a massless scalar $\chi$. Their quadratic actions have the shift symmetries
	\begin{align}\label{eq:shiftsymmetry}
		\delta A_i = \xi^\text{KV}_i \;, \qquad \delta \phi^a =  { \nabla^i} \xi_i^{\text{CKV},a} \;, \qquad \delta \chi = \text{constant}\;.
	\end{align}	
	Here (C)KV denotes (conformal) Killing vectors and $\nabla_i$ the covariant derivative on the round $S^{D-2}$ with radius $\ell$. The shift symmetries indicate that $A_i$, $\phi^a$ and $\chi$ may be Goldstones of some underlying symmetry breaking. The bulk-edge factorization \eqref{eq:Z1loopbulkedge} suggests a complementary viewpoint: the sphere path integral encodes the thermal bulk physics of a particular static patch, ``dressed" by a horizon contribution within a diffeomorphism-invariant gravitational partition function.\footnote{Symmetry-breaking perspectives on static-patch observers have also been suggested in \cite{Boyda:2002ba,Anninos:2011af,Susskind:2025wyn,Susskind:2026fjc,Susskind:2026moe}.} 
	
	Further clues to the geometric nature of these modes come from work on subsystems in gauge theories and gravity.\footnote{For further work on edge modes and related boundary degrees of freedom, see, e.g., \cite{Kabat:1995eq,Donnelly:2011hn,Donnelly:2012st,Casini:2013rba,Radicevic:2014kqa,Donnelly:2014gva,Donnelly:2014fua,Ghosh:2015iwa,Donnelly:2015hxa,Soni:2015yga,Delcamp:2016eya,Agarwal:2016cir,Wong:2017pdm,Blommaert:2018oue,Barnich:2018zdg,Freidel:2018fsk,Gomes:2018dxs,Donnelly:2018ppr,Gomes:2019xto,Geiller:2019bti,Riello:2021lfl,Mukherjee:2023ihb,He:2024skc,fewster2025semi} for gauge theories and  \cite{Balachandran:1994EdgeStates,Geiller:2017xad,Geiller:2017whh,Freidel:2019ees,Benedetti:2019uej,Takayanagi:2019tvn,Freidel:2020svx,Freidel:2020ayo,Donnelly:2020xgu,Freidel:2021cjp,Ciambelli:2021nmv,Freidel:2021dxw,Anninos:2021ihe,David:2022jfd,Ciambelli:2022cfr,Mertens:2022ujr,Wong:2022eiu,Donnelly:2022kfs,Balasubramanian:2023dpj,Ciambelli:2024qgi,blommaert2025gravitons,Langenscheidt:2024nyw,Joung:2025Extracting,Freidel:2025AnomalousFrames,Ciambelli:2025ztm,Ciambelli:2026pwi} for gravity. For a review of gravitational corner symmetries, see \cite{Ciambelli:2022vot}. For related work on entanglement and edge modes in string theory, see, e.g., \cite{Dabholkar:1994ai,Donnelly:2016jet,Donnelly:2020teo,Jiang:2020cqo,Dabholkar:2023ows,Dabholkar:2023tzd,Dabholkar:2024neq,Dabholkar:2026afn,Wong:2026ftn}. }  First, for gauge theories in static spacetimes, edge modes have been  understood as large gauge transformations supported on the stretched horizon \cite{Blommaert:2018rsf,Ball:2024hqe,Ball:2024xhf,Ball:2024gti}. In gravity, corner phase-space analyses similarly identify edge degrees of freedom associated with large diffeomorphisms  \cite{Donnelly:2016auv,Speranza:2017gxd,Freidel:2020xyx,Ciambelli:2021vnn}. Second, work on dynamical reference frames \cite{Carrozza:2021gju,Carrozza:2022xut,Kabel:2023BoundaryFrames,Giesel:2024EdgeClocks,Araujo-Regado:2024dpr,Araujo-Regado:2025ejs} and von Neumann algebras \cite{Chandrasekaran:2022cip,DeVuyst:2024khu,DeVuyst:2024fxc,Chandrasekaran:2026pnc,Klinger:2026tws} has connected edge modes with observers.
	
	In this work, we bring these perspectives together to make the symmetry-breaking interpretation of de Sitter horizon edge modes precise. We focus on the case of $D\geq 4$. As we will see, the modes \eqref{eq:shiftsymmetry} can be understood in terms of diffeomorphism jets near the horizon sphere and as Goldstones associated with symmetry breaking induced by specifying a {\it test} observer in $dS_D$, specified by the framed embedding of their horizon $S^{D-2}$. With this geometric description, we will construct nonlinear effective field theories (EFTs) that reproduce the spectra \eqref{eq:Zedgegravresult}. 
	
	\section{$dS_D$ horizon edge modes from diffeomorphism jets}\label{sec:edgediff}
		
		In this section, we identify the geometric origin of \eqref{eq:shiftsymmetry}. We start with a static patch in de Sitter space $dS_{{ D}}$ with radius $\ell${,}
	\begin{align}\label{eq:tubularcoor}
		ds_0^2
		&= \eta_{ab}d\sigma^a d\sigma^b
		+\frac{(\sigma_a d\sigma^a)^2}{\ell^2-\sigma^2}
		+ \left(1-\frac{\sigma^2}{\ell^2}\right)\gamma_{ij}(x)dx^i dx^j\; ,
	\end{align}
	where $\eta_{ab}=\mathrm{diag}(-,+)$ with $a,b=0,{ D}$, and $\sigma^2\coloneqq\eta_{ab}\sigma^a\sigma^b$. The metric $\gamma_{ij}\coloneqq \ell^2\delta_{IJ}\partial_i\Omega^I \partial_j\Omega^J$, $ I,J=1,\ldots,{ D-1}$, where $\delta_{IJ}\Omega^I \Omega^J=1$, is the round metric on $S^{{ D-2}}$ of radius $\ell$, with $i,j=1, \ldots,{ D-2}$ denoting local $S^{{ D-2}}$ coordinates.
	The coordinates $\sigma^a,x^i$ have dimensions of length; $\Omega^I,\chi$ and the metric components $\gamma_{ij}$ are dimensionless.
	
	The coordinates $\sigma^a\coloneqq(\sigma^0,\sigma^{{ D}})$ are related to the familiar static coordinates through
	\begin{align}
		\sigma^0 = \sqrt{\ell^2-r^2} \, \sinh \frac{t}{\ell}\; ,
		\qquad
		\sigma^D =\sqrt{\ell^2-r^2} \, \cosh \frac{t}{\ell}\; ,
	\end{align}
	so that the bifurcation sphere $S^{ { D-2}}$ sits at $\sigma^a=(0,0)$.
	
	The spacetime $dS_D$ has $\frac{D(D+1)}{2}$ Killing vectors, denoted by $L_{\mathsf A\mathsf B}=-L_{\mathsf B\mathsf A}=L_{\mathsf A\mathsf B}^\mu\partial_\mu$, $\mathsf A,\mathsf B=0,1,\dots,D$. The Killing vectors that preserve the region bounded by the static patch horizon are the $SO({ D-1})$ rotations, $L_{\mathsf I\mathsf J}$, and the de Sitter boost, or static patch Hamiltonian, $L_{0,{ D}}$,
	\begin{equation}\label{eq:dS4staticKilling}
		{ L_{\mathsf I\mathsf J} = \mathcal{J}_{\mathsf I\mathsf J}^{\,i}(x)\partial_i} \;, \qquad L_{0, { D}} =-\sigma^0 \partial_{\sigma^{{ D}}}-\sigma^{{ D}} \partial_{\sigma^0}=-\ell\, \partial_t \; .
	\end{equation}
	Here ${\mathcal{J}_{\mathsf I\mathsf J}^{\,i}(x)}$ denotes the ${\frac{(D-1)(D-2)}{2}}$ Killing vectors on a unit round $S^{{ D-2}}$. The ${ 2(D-1)}$ remaining Killing vectors 
	\begin{align}\label{eq:dS4brokenKillingExact}
		L_{\mathsf a\mathsf I} &
		 = \delta_{\mathsf a}^{a}\delta_{\mathsf I}^{I}\left[\ell^2\frac{\sigma_a}{\sqrt{\ell^2-\sigma^2}}\,\gamma^{ij}\partial_j \Omega_I\,\partial_i-\sqrt{\ell^2-\sigma^2}\,\Omega_I\,\partial_{\sigma^a}\right] = -\ell\,\delta_{\mathsf a}^{a}\delta_{\mathsf I}^{I}\Omega_I\,\partial_{\sigma^a}+O(\sigma)\;,
	\end{align}
move points in or out of the static patch. We have expanded near $\sigma^a = (0,0)$ in the second equality.

Now, consider a diffeomorphism $\Phi : dS_D \to dS_D$ and expand it around 
$\sigma^a=(0,0)$,
\begin{align}\label{eq:fulljetdata}
	\Phi^\mu(\sigma,x)&=\Phi_0^\mu(x)+\sigma^aJ_a{}^\mu(x)+O(\sigma^2) \;.
\end{align}
Here $\Phi_0^\mu(x)\coloneqq\Phi^\mu(0,x)$ and $J_a{}^\mu(x)\coloneqq\partial_a\Phi^\mu(0,x)$ are the 0- and transverse 1-jets, respectively.

Let $v$ be a Killing vector generating a finite isometry $g_v$ of $dS_D$. As a diffeomorphism, this generates an action on $\Phi$ by group composition,
\begin{align}\label{eq:diffleftaction}
	\Phi &\mapsto g_v\circ \Phi \;, \qquad  g_v^\mu(x)\coloneqq e^{v^\nu(x)\partial_\nu}x^\mu \; ,
\end{align}
or, infinitesimally,
\begin{align}\label{eq:diffnonlineartrans}
	\delta_v \Phi ^\mu(\sigma,x)
	&=
	v^\mu\bigl(\Phi(\sigma,x)\bigr) \;.
\end{align}
This transformation is infinitesimal in $v$ but fully nonlinear in $\Phi$. To identify the edge mode fields \eqref{eq:shiftsymmetry}, we expand the diffeomorphism \eqref{eq:fulljetdata} around the identity map, i.e.,
\begin{align}\label{eq:diffexpand}
	\Phi^\mu (\sigma,x)&= \left( \sigma^a,x^i\right)  + \xi^\mu (\sigma,x) + O(\xi^2){,}
\end{align}
with small $\xi^\mu$. Here $O(\xi^2)$ are terms that depend on the specific nonlinear completion. For example, $\Phi_0^i=x^i+\xi^i$ and $\Phi_0^i=\left( e^{\xi^j\partial_j}x\right) ^i$ satisfy \eqref{eq:diffexpand} but differ at higher orders. These different choices are related by a nonlinear field redefinition. {Eq.} \eqref{eq:diffexpand} implies the expansion of the jets  
\begin{gather}
	\Phi_0^a(x)=\xi^a(0,x)+O(\xi^2)\;,\qquad
	\Phi_0^i(x)=x^i+\xi^i(0,x)+O(\xi^2)\;,\nn\\
	J_a{}^\mu(x)=\delta_a^\mu+\partial_a\xi^\mu(0,x)+O(\xi^2)\;.\label{eq:lowjets}
\end{gather}

Now, the key observation is that the transformation \eqref{eq:diffnonlineartrans} implies the following non-trivial $O(\xi^0)$-transformation laws of the jet fluctuations \eqref{eq:lowjets} under the $dS_D$ isometries \eqref{eq:dS4staticKilling} and \eqref{eq:dS4brokenKillingExact}:
\begin{gather}
	{\delta_{\mathsf I\mathsf J}} \xi^i\big|_{\sigma=0}  = {\mathcal{J}_{\mathsf I\mathsf J}^{\,i}}  +
	O\left( \xi^1\right) \;, \qquad 
	{\delta_{\mathsf a\mathsf I}}\left(\xi^b\big|_{\sigma=0}\right)= -\ell\,{\delta_{\mathsf I}^{I}\Omega_I \delta^b_{\mathsf a}} +
	O\left( \xi^1\right) \;, \nn\\
	\delta_{0, { D}} \left( \frac12\epsilon^{(\perp)}_{ab}\partial^a	\xi^b\big|_{\sigma=0}\right) =1+
	O\left( \xi^1\right) \;,\qquad {\delta_{\mathsf a\mathsf I}}\left(\left.\partial^b\xi^i\right|_{\sigma=0}\right)
	=
	\ell\,{\delta^b_{\mathsf a}\delta_{\mathsf I}^{I}} { \nabla^i}\Omega_I +
	O\left( \xi^1\right) \;. \label{eq:jettransform}
\end{gather}
Here $\delta_{\mathsf A\mathsf B}\coloneqq\delta_{L_{\mathsf A\mathsf B}}$, $\nabla_i$ is the {Levi-Civita} connection on the round sphere $S^{{ D-2}}$ of radius $\ell$, and $\epsilon^{(\perp)}_{0,{ D}}=+1$, with transverse indices raised by $\eta^{ab}$. Since $\mathcal{J}_{\mathsf I\mathsf J}^{\,i}$ are round $S^{D-2}$ Killing vectors,
while $\nabla^i\Omega_I$ are conformal Killing vectors with divergence
$\nabla^2\Omega_I=-(D-2)\Omega_I/\ell^2$, the first three transformation laws in
\eqref{eq:jettransform} reproduce the shifts \eqref{eq:shiftsymmetry}.
We therefore identify
\begin{align}\label{eq:edgemodejetiden}
	A^i&=\xi^i(0,x)+O(\xi^2)\;,\qquad
	\phi^a=\xi^a(0,x)+O(\xi^2)\;,\qquad 
	\chi=\frac12\epsilon^{(\perp)}_{ab}\partial^a\xi^b(\sigma,x)\big|_{\sigma=0}+O(\xi^2)\;.
\end{align}
Henceforth we refer to these jets as ``edge modes".\footnote{The term ``edge modes'' has several  uses in the literature that need not coincide with our usage here, including Stueckelberg-like boundary fields introduced in extended phase-space constructions \cite{Donnelly:2016auv,Speranza:2017gxd}.}  The shift transformations \eqref{eq:jettransform} suggest an underlying symmetry breaking phenomenon, which we will establish in Section \ref{sec:staticpatchembedding}.

	To end this discussion, suppose we have a diffeomorphism $\Phi$ containing the jet fluctuations \eqref{eq:edgemodejetiden}. A family of metrics immediately follows by pulling back the metric \eqref{eq:tubularcoor} under $\Phi${,}
	\begin{align}\label{eq:largeDiffEdgeMetric}
		ds_\Phi^2 \coloneqq \left( g_\Phi\right) _{\mu\nu}dx^\mu dx^\nu 
		\coloneqq\Phi^*ds_0^2
		= \eta_{ab}d\Phi^a d\Phi^b
		+\frac{\left( \Phi_a d\Phi^a\right) ^2}{\ell^2-\Phi^2}
		+ \left( 1-\frac{\Phi^2}{\ell^2}\right) \gamma_{ij}(\Phi)d\Phi^i d\Phi^j\;.
	\end{align}
	Here $\Phi^2\coloneqq\eta_{ab}\Phi^a\Phi^b$. For any diffeomorphism-invariant theory admitting a $dS_D$ background, this (trivially) corresponds to a family of {\it nonlinear} solutions to the classical equation of motion. Note that the {horizon sphere} is no longer at $\sigma^a=0$ but instead at $\Phi^a(\sigma,x)=0$.
	
	On the full $dS_D$ spacetime, the metrics \eqref{eq:largeDiffEdgeMetric} are all gauge-equivalent. The question here is whether they remain gauge-equivalent when restricted to a static-patch subregion. Relatedly, why are the 0- and 1-jets \eqref{eq:edgemodejetiden} singled out from higher jets and the rest of the 1-jets transforming as the last equation in \eqref{eq:jettransform}? In {Appendix}~\ref{app:CPS}, we perform a preliminary covariant phase space analysis in Einstein gravity and show how the jets \eqref{eq:edgemodejetiden} correspond to nondegenerate phase space directions, with boundary conditions and corner ambiguities left for future analysis. We will also give another perspective on the restriction of the 1-jets to \eqref{eq:edgemodejetiden} in {Section} \ref{sec:EFT}.

\section{Observer frames, symmetry breaking and Goldstone modes}\label{sec:staticpatchembedding}

In this section, we show that the edge modes \eqref{eq:edgemodejetiden} are {Goldstone modes} associated with symmetry breaking induced by specifying a  test observer in de Sitter space. To that end, we consider the geometric data that characterizes {\it a} static patch such as \eqref{eq:tubularcoor}. 

First, the horizon sphere at $\sigma^a=(0,0)$ can be invariantly viewed as an isometric, totally geodesic embedding of a fixed round reference sphere, $S^{{ D-2}}_{\rm int}${,} of radius $\ell$,
\begin{align}\label{eq:embedmap}
	\iota_0:S^{{ D-2}}_{\rm int}\to dS_{{ D}}\; .
\end{align}
Here ``isometric'' means $\iota_0^*g_0=\gamma$, where $\gamma$ is the round metric on $S^{ { D-2}}_{\rm int}$. In a de Sitter space,  the map \eqref{eq:embedmap} fixes the bifurcation sphere, which picks out a pair of complementary causal  regions. 

\begin{figure}[H]
\centering
\includegraphics[width=\textwidth]{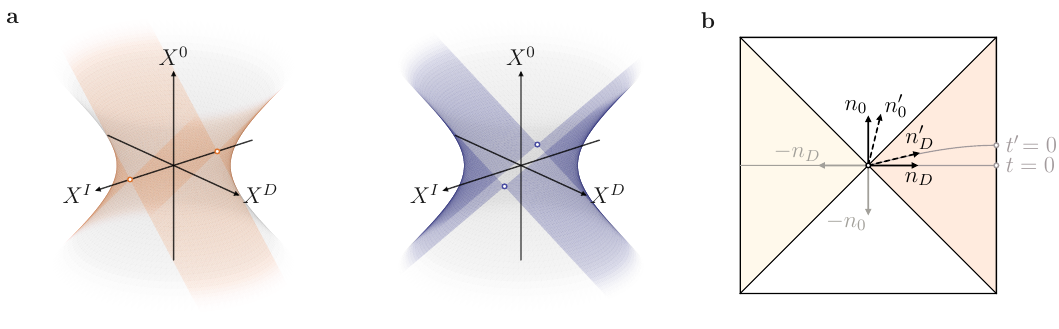}
\caption{(a) Different choices of bifurcation spheres (orange and blue) pick out different pairs of complementary causal regions. (b) For a fixed bifurcation sphere, different choices of normal frames $n_{\mathsf a}$ or $n'_{\mathsf a}$ (related by an $SO(1,1)$ boost) change the preferred spatial slice. The opposite arrows $-n_{\mathsf 0}$ and $-n_{{ D}}$ respectively reverse the time orientation and exchange the complementary regions.}
\label{fig:static-patch-data-choices}
\end{figure}

To fully specify a patch \eqref{eq:tubularcoor}, we equip the horizon sphere with a parallel normal frame, which consists of a pair of orthonormal vector fields $\left( n_0^{(0)},n_{{ D}}^{(0)}\right) $ orthogonal to all tangent vectors pointwise on the embedded sphere\footnote{For a totally geodesic surface in $dS_{{ D}}$ with $D\geq4$, the Gauss equation gives a constant sectional curvature $1/\ell^2$, so the image is a round $S^{{ D-2}}$ of radius $\ell$. The Ricci equation implies a flat normal-bundle connection; since $S^{{ D-2}}_{\rm int}$ is simply connected, this connection has trivial holonomy and admits a global parallel orthonormal normal frame, unique up to a constant $SO(1,1)$ transformation and discrete orientation choices.}
\begin{align}\label{eq:normalframe}
	g_0(n_{\mathsf a}^{(0)},\iota_{0*}u)=0\quad \left( u\in TS^{{ D-2}}_{\rm int}\right) \; ,\quad
	g_0(n_{\mathsf a}^{(0)},n_{\mathsf b}^{(0)})=\eta_{\mathsf a\mathsf b}\; ,\quad \nabla^\perp_i n_{\mathsf a}^{(0)} =0\;, \quad \mathsf a,\mathsf b=0,{ D}\; .
\end{align}
Here $g_0(\cdot,\cdot)$ is the bilinear form defined by the metric \eqref{eq:tubularcoor}. We use $\mathsf a, \mathsf b$ for normal-frame indices and $a,b$ for transverse coordinate indices. The normal-bundle covariant derivative for a normal vector field $V$ along $\iota_0(S^{ { D-2}}_\text{int})$ is defined as the normal projection of the Levi-Civita derivative $\nabla^{dS}$ associated with $g_0$, with $\Pi_\perp$ denoting the orthogonal projector:
\begin{align}\label{eq:normalbundlecovariantderivative}
	\nabla_i^\perp V
	& \coloneqq \Pi_\perp\nabla^{dS}_iV\;,\qquad
	\Pi_\perp W\coloneqq n_{\mathsf a}^{(0)}\eta^{\mathsf a\mathsf b}g_0\left(n_{\mathsf b}^{(0)},W\right)\;,\qquad
	W\in TdS_{{ D}}\big|_{\iota_0\left( S^{{ D-2}}_\text{int}\right) }\;.
\end{align}
The spacelike normal $n_{D}^{(0)}$ selects a preferred ``inward" radial direction, which also generates a distinguished $t=0$ slice. Meanwhile, the timelike normal  $n_0^{(0)}$ fixes the time orientation, i.e., which direction of the associated static time flow is future-directed.\footnote{While the coordinates \eqref{eq:tubularcoor} are adapted to {\it this} framed embedding,
	\begin{align}\label{eq:adaptedcoor}
		{(\iota_0)}^\mu(x)=(\sigma^a=0,x^i)\;, \qquad n_{\mathsf a}^{(0)}=\left.\frac{\partial}{\partial \sigma^a}\right|_{\sigma=0}\;, 
		\qquad \mathsf a=0,{ D}\; ,
	\end{align}
	we stress that the pair $\left(\iota_0,n_{\mathsf a}^{(0)}\right)$ of geometric data is not coordinate-dependent.} See {Figure} \ref{fig:static-patch-data-choices}. 

In perturbative gravity, the pair $(\iota_0,n^{(0)}_{\mathsf a})$ can be viewed as an {\it intrinsic} notion of an observer, defined as a reference geometric  datum of the background $dS_D$ geometry, relative to which gravitational perturbations are studied. Such a {\it test} observer carries no stress tensor or other charges and therefore does not backreact on the geometry, in contrast to a {\it physical} observer modeled by a backreacting object such as a worldline \cite{Chandrasekaran:2022cip} or by an additional dynamical field \cite{Chen:2024rpx}.
	
	Now, we consider the space of all such test-observer data in $dS_D$:
		\begin{align}\label{eq:totalspace}
			\mathcal{C}
			= \left\{(\iota,n_{\mathsf a}):
			\begin{array}{l}
				\iota: S^{{ D-2}}_\text{int} \to dS_{{ D}} \text{ isometric, totally geodesic embedding}\;,\\
				n_{\mathsf a} \text{ ordered orthonormal normal frame along } \iota \left( S^{{ D-2}}_\text{int}\right)\; ,
					\quad \nabla_i^\perp n_{\mathsf a}=0
			\end{array}
			\right\} \;.
		\end{align}
The space \eqref{eq:totalspace} has a natural symmetry group\footnote{Here we restrict to the orientation-preserving subgroup $SO(1,{ D})_\text{ext}\times SO({ D-1})_\text{int}$. One could also consider the larger group $O(1,{ D})_\text{ext}\times O({ D-1})_\text{int}$, which includes orientation-changing transformations such as $n_0\to-n_0$ with $n_D$ fixed.}
\begin{align}\label{eq:fullsymmetry}
	SO(1,{ D})_\text{ext}\times SO({ D-1})_\text{int}\;, 
\end{align}
which acts on any pair $(\iota,n_{\mathsf a})\in \mathcal{C}$ by 
\begin{align}\label{eq:framedembedsym}
    \big((L,\bar R)\cdot\iota\big)(p)
    =L\big(\iota(\bar R^{-1}p)\big)\; ,\qquad 
	\big((L,\bar R)\cdot n\big)_{\mathsf a}(p)
	=L_*\,n_{\mathsf a}(\bar R^{-1}p)\; ,
\end{align}
with $p\in S^{{ D-2}}_\text{int} $ and $\left( L,\bar R\right) \in SO(1,{ D})_\text{ext} \times SO({ D-1})_\text{int}$. {Here $L_*$ denotes the pushforward by $L$.}

For any chosen framed embedding \((\iota,n_{\mathsf a})\in\mathcal C\), there is a distinguished rotation subgroup $SO({ D-1})_\iota\subset SO(1,{ D})_{\rm ext}$, identified with \(SO({ D-1})_{\rm int}\) through equivariance conditions
	\begin{align}\label{eq:equivariance}
		\iota(\bar R p)=R\,\iota(p),\qquad
		n_{\mathsf a}(\bar R p)=R_*n_{\mathsf a}(p)\; .
	\end{align}
Here $R_*$ denotes the pushforward by $R\in SO({ D-1})_{\iota}$. The conditions \eqref{eq:equivariance} simply mean that rotating a point $p\in S^{{ D-2}}_\text{int}$ by $\bar R\in SO({ D-1})_\text{int}$ and then embedding gives the same target point as embedding first and then applying the corresponding target rotation $R\in SO({ D-1})_{\iota}$. Together with \eqref{eq:framedembedsym}, we see that a generic element of \eqref{eq:fullsymmetry} moves $(\iota,n_{\mathsf a})$ to a different framed embedding $(\iota',n'_{\mathsf a})$, while the diagonal subgroup
\begin{align}\label{eq:diagonalSO3}
SO({ D-1})_{\rm diag}
=
\{(R,\bar R):\bar R\in SO({ D-1})_{\rm int},\ R\in SO({ D-1})_\iota
\text{ corresponds to }\bar R\}
\end{align}
leaves it invariant. To conclude, the choice of {\it a} static patch, characterized by the background observer datum $(\iota,n_{\mathsf a})\in \mathcal{C}$, breaks the full symmetry \eqref{eq:fullsymmetry} into the diagonal subgroup \eqref{eq:diagonalSO3}:
	\begin{align}\label{eq:edgesymbreak}
			SO(1,{ D})_\text{ext}\times SO({ D-1})_\text{int}\to SO({ D-1})_\text{diag} \;.
	\end{align}

We now return to the full diffeomorphism jets $(\Phi_0^\mu,J_a{}^\mu)$ in \eqref{eq:fulljetdata}, without restricting to the fluctuations \eqref{eq:edgemodejetiden}. We first implement the full symmetry \eqref{eq:fullsymmetry} by extending the left action \eqref{eq:diffleftaction} to the product group 
	\begin{align}\label{eq:fulljetmapaction}
		\left( (L,\bar R)\cdot\Phi\right) (\sigma,x)
		=L\big(\Phi(\sigma,\bar R^{-1}x)\big)\; ,\qquad
		(L,\bar R)\in SO(1,{ D})_\text{ext}\times SO({ D-1})_\text{int}\; .
	\end{align}
	Expanding around $\sigma=0$ implies the finite group action on the 0- and 1-jets
	\begin{align}\label{eq:fulljetfiniteaction}
		\Phi_0'{}^\mu(x)=L^\mu\bigl(\Phi_0\left(\bar R^{-1}x\right)\bigr)\;,\qquad 
		J'_a{}^\mu(x)&=\partial_\nu L^\mu\bigl(\Phi_0\left(\bar R^{-1}x\right)\bigr)J_a{}^\nu\left(\bar R^{-1}x\right)\;.
	\end{align}
These transformations depend only on the original 0- and 1-jets, so this space is closed under the full symmetry \eqref{eq:fullsymmetry}.\footnote{In general, for every finite $k$, the space of all $(\leq k)$-jets is preserved by \eqref{eq:fulljetmapaction}.} At the identity map, these jets describe the reference configuration $(\iota_0,n_{\mathsf a}^{(0)})$, which is preserved by $SO({ D-1})_\text{diag}$. The broken symmetries act nonlinearly on the jets and move this reference configuration, reproducing the shifts \eqref{eq:jettransform} at leading order in the fluctuations. Therefore, the gravitational edge modes \eqref{eq:edgemodejetiden} can be viewed as Goldstone fields for the symmetry breaking pattern \eqref{eq:edgesymbreak}.

Geometrically, the full 0-jet $\Phi_0^\mu(x)$ is the coordinate expression of the deformed embedding $\iota_\Phi\coloneqq\Phi\circ\iota_0$, which need not remain round or totally geodesic, while the 1-jet determines the pushforward of the reference normal frame \eqref{eq:adaptedcoor}:
\begin{align}\label{eq:jetFramePushforward}
    \left.\Phi_*n_{\mathsf a}^{(0)}\right|_{\Phi_0(x)}
    &= \left.\frac{\partial\Phi^\mu}{\partial\sigma^a}\right|_{(0,x)}\left.\partial_\mu\right|_{\Phi_0(x)}
    =J_a{}^\mu(x)\left.\partial_\mu\right|_{\Phi_0(x)}\;,
\end{align}
which need not be orthogonal to the deformed surface or orthonormal with respect to the background metric \eqref{eq:tubularcoor}.

In {Section} \ref{sec:EFT}, we construct the adapted orthonormal normal frame and the induced geometric quantities that become the building blocks of the effective action, and address the remaining 1-jet components. In {Section} \ref{sec:perturbative}, we explain how this geometry relates to the fields $A_i$, $\phi^a$, and $\chi$ in \eqref{eq:edgemodejetiden}.

	\section{Effective field theory of de Sitter horizon edge modes}\label{sec:EFT}
			
		We now construct {EFTs} governing the edge modes \eqref{eq:edgemodejetiden}, based on the  geometric picture explained in Section \ref{sec:staticpatchembedding}. As Goldstones for the symmetry breaking pattern \eqref{eq:edgesymbreak}, their dynamics is heavily constrained by the symmetries \eqref{eq:fullsymmetry}.  In Appendix~\ref{sec:coset}, we apply an alternative  method known as the coset construction \cite{Coleman:1969sm,Callan:1969sn,Volkov:1973vd,Ogievetsky1974}, with emphasis on the so-called inverse Higgs constraints (IHCs), which we show recover an equivalent set of building blocks for the EFT.

We focus on the static patch \eqref{eq:tubularcoor} with round horizon embedding $\iota_0$. A diffeomorphism $\Phi$ containing the jets \eqref{eq:fulljetdata} deforms the reference embedding to $\iota_\Phi=\Phi\circ\iota_0$ and pulls back the background metric to $\Phi^*g_0$, as summarized by the following commutative diagram:
	\begin{align}
			\begin{tikzcd}[ampersand replacement=\&,row sep=2.2em,column sep=4.5em]
				\text{internal sphere }\left(S^{{ D-2}}_\text{int},\iota_\Phi^*g_0\right)
				\arrow[r,"\iota_0"] \arrow[dr,"\iota_\Phi"']
				\& \left(dS_{{ D}},\Phi^*g_0\right)\arrow[d,"\Phi"] \text{ pullback geometry}\\
				\& \left(dS_{{ D}},g_0\right) \text{ background}
			\end{tikzcd}
	\end{align}
	While the internal $S^{{ D-2}}_\text{int}$ is already equipped with a fixed round metric $\gamma_{ij}$, these maps induce new geometric quantities, which become the building blocks for the effective action.

The simplest geometric quantity for the embedded surface is the induced metric
	\begin{align}\label{eq:generalJetInducedMetric}
		h_{ij}^{(\Phi)}
		\coloneqq\left(\iota_\Phi^*g_0\right)_{ij}= \left(\eta_{ab}
			+\frac{\Phi_{0a}\Phi_{0b}}{\ell^2-\Phi_0^2}\right)
			\partial_i\Phi_0^a\,\partial_j\Phi_0^b+\left(1-\frac{\Phi_0^2}{\ell^2}\right)
			\gamma_{kl}(y)\,\partial_i y^k\,\partial_j y^l
	\end{align}
	 on $S^{{ D-2}}_\text{int}$. Here $\Phi_0^2\coloneqq\eta_{ab}\Phi^a\Phi^b\big|_{\sigma =0}$ and $	y^i\coloneqq\left.\Phi^i\right|_{\sigma =0}$. 
	
 	The effective action includes all invariants constructed using either the fixed round metric $\gamma_{ij}$ or the induced metric $h^{(\Phi)}_{ij}$ for contractions and covariant derivatives. Equivalently, we can express the latter in terms of the tensor
	 	\begin{align}\label{eq:straintensor}
	 		\mathsf S ^i{}_j[\Phi]\coloneqq\left( \sqrt{\gamma^{-1}h^{(\Phi)}}\right) ^i{}_j-\delta ^i{}_j \;,
	 	\end{align}
	 	which measures the deviation from the round metric $\gamma_{ij}$. In analogy with continuum mechanics, we refer to \eqref{eq:straintensor} as the Biot strain tensor \cite{Anand:2020cms}. For example, the inverse metric is $ \left( h^{(\Phi)}\right)^{-1}=\left(I+\mathsf S\right)^{-2}\gamma^{-1}$.  Likewise, covariant derivatives compatible with $h^{(\Phi)}$ can be rewritten using the connection difference $C^k{}_{ij}\left[ \Phi\right] 
	 	\coloneqq \Gamma[h^{(\Phi)}]^k{}_{ij}-\Gamma[\gamma]^k{}_{ij}$, which only depends on the Biot strain \eqref{eq:straintensor} and its first derivatives. Finally, the induced volume measure is $\sqrt{\det h^{(\Phi)}}=\sqrt{\gamma}\,\det(I+\mathsf S[\Phi])$.

	Other basic geometric quantities of an embedded surface are its extrinsic curvature and normal-bundle connection, which are defined with respect to an adapted orthonormal normal frame. However, the pushforward $\Phi_*n_{\mathsf a}^{(0)}$ in \eqref{eq:jetFramePushforward} is generally neither normal to the deformed sphere $\iota_\Phi\left( S^{{ D-2}}_\text{int}\right) $  nor orthonormal. This motivates us to impose\footnote{In the coset construction in {Appendix} \ref{sec:coset}, orthogonality arises as one of the inverse Higgs constraints \eqref{eq:IHC1}.}
	\begin{align}\label{eq:invarconstraint}
		(g_\Phi)_{ai}|_{\sigma=0}=0\;,\qquad
		(g_\Phi)_{ab}|_{\sigma=0}= \eta_{ab}\;,
	\end{align}
	which requires $\Phi_*n_{\mathsf a}^{(0)}$ to be orthogonal to $\iota_\Phi\left( S^{{ D-2}}_\text{int}\right) $ and orthonormal. Every locally framed spacelike embedding can be represented by jets satisfying these constraints.

	The constraints \eqref{eq:invarconstraint} are invariant under the full symmetry \eqref{eq:fulljetmapaction}, so the constrained space of 0- and 1-jets remains closed. At fixed embedding $\Phi_0^\mu$, the orthogonality and normalization conditions provide respectively $2(D-2)$ and 3 constraints that express $2D-1$ of the $2D$ transverse first-jet components $J_a{}^\mu$ in terms of the remaining 0- and 1-jets. The remaining freedom is an $SO(1,1)$ boost of the normal frame, which preserves \eqref{eq:invarconstraint}. Together with the embedding fields, this gives nonlinear extensions of $A_i,\phi^a,\chi$ in \eqref{eq:edgemodejetiden}.

	We denote the adapted orthonormal normal frame by $n_{\mathsf a}^{(\Phi)}\coloneqq \Phi_*n_{\mathsf a}^{(0)}$ with \eqref{eq:invarconstraint} imposed, and the tangent vectors by $t_i^{(\Phi)}\coloneqq(\iota_\Phi)_*\partial_i$. The extrinsic curvature is
	\begin{align}\label{eq:extrinsiccurvature}
		K^{\mathsf a}{}_{ij}[\Phi]
		&\coloneqq g_0\!\left(n^{(\Phi)\mathsf a},
		\left(\iota_\Phi^*\nabla^{dS}\right)_i t_j^{(\Phi)}\right)
		=K^{\mathsf a}{}_{ji}\;,
	\end{align}
	and the normal-bundle connection is
	\begin{align}\label{eq:normalbundleconnection}
		{\mathcal A_i}[\Phi]
		&\coloneqq\frac12\epsilon^{(\perp)\mathsf a\mathsf b}
		g_0\!\left(n_{\mathsf a}^{(\Phi)},
		\left(\iota_\Phi^*\nabla^{dS}\right)_i n_{\mathsf b}^{(\Phi)}\right)\;.
	\end{align}
	Notice that while geometrically \eqref{eq:normalbundleconnection} is a connection, it is invariant under the nonlinearly realized $SO(1,{ D})_\text{ext}$ and transforms as a 1-form under the linearly realized $SO({ D-1})_\text{diag}$, and thus should be included as a building block for the effective action. The blocks \eqref{eq:straintensor}, \eqref{eq:extrinsiccurvature}, \eqref{eq:normalbundleconnection}, together with their derivatives and the fixed background data, determine the local intrinsic and extrinsic geometry of the framed embedding.
	
The {resulting} effective action realizing the symmetry breaking pattern \eqref{eq:edgesymbreak} is constructed from \eqref{eq:straintensor}, \eqref{eq:extrinsiccurvature}, \eqref{eq:normalbundleconnection}, their $\gamma$-covariant derivatives, together with their $\gamma$-contractions:
\begin{align}\label{eq:generalEFT}
	I=\int_{S^{D-2}}\, F\left[ 	\gamma_{ij}\,, { \nabla_i}\, , \mathsf S^i{}_j\left[ \Phi\right]  \,, K^{\mathsf a}{}_{ij}\left[ \Phi\right] \,,   {\mathcal A_i}\left[ \Phi\right] \right] \;,
\end{align}
where $F$ is a local scalar built from the listed building blocks. We restrict to strictly invariant local densities and leave possible Wess-Zumino terms for future investigation. We use the shorthand $\int_{S^{D-2}}d^{{ D-2}}x \sqrt{\gamma}\to \int_{S^{D-2}}$ for the fixed volume measure on the round $S^{D-2}$ of radius $\ell$.

 In Einstein gravity, the Planck mass $M_{\mathrm{Pl}}$ sets the suppression scale of higher-derivative operators.  By dimensional analysis, the building blocks carry the following powers of $\ell$, 	
 \begin{align}\label{eq:geometricGradientCounting}
		\mathsf S^i{}_j\propto\ell^0 \;,\qquad
		K^{\mathsf a}{}_{ij}\;, \mathcal A_i\; ,  |{ \nabla_i}| \propto \ell^{-1}\;,
	\end{align}
	up to dimensionless factors. Assuming these dimensionless factors remain $O(1)$ as $\ell$ grows, we have a natural organization of \eqref{eq:generalEFT}  as an expansion in inverse powers of $M_\text{Pl}\ell\gg 1$,\footnote{For more general gravitational theories, there could be some intermediate scale $L$ with the hierarchy $M_\text{Pl}^{-1}\ll L\ll\ell$, in which case the perturbative expansion \eqref{eq:geometricGradientAction} should be organized by the parameter  $L/\ell$ instead. Here we simply assume $M_\text{Pl}$ is the only UV scale.} 
	\begin{align}\label{eq:geometricGradientAction}
		I=\sum_{N\geq0}\left( \frac{1}{M_\text{Pl}\ell}\right)^N I^{(N)}\left[\mathsf S,\ell K,\ell\mathcal A;\ell {\nabla},\ell^2R\right]\;,
		\qquad 
		N=n_K+n_{\mathcal A}+n_{\nabla}+2n_R\;.
	\end{align}
	In each term  $I^{(N)}$, $n_K,n_{\mathcal A},n_{\nabla},n_R$ count factors of $K$, $\mathcal A$, additional covariant derivatives, and curvature $R_{ijkm}=(\gamma_{ik}\gamma_{jm}-\gamma_{im}\gamma_{jk})/\ell^2$, respectively. Since $\mathsf S$ carries no inverse powers of $\ell$, its arbitrary powers can appear at every order. In particular, the lowest order action is
	\begin{align}\label{eq:leadingStrainAction}
		I^{(0)}=\frac{1}{8\pi G_N}\int_{S^{{ D-2}}}\, B(\mathsf S){,}
	\end{align}
	where  $B$ is a dimensionless scalar function of $\mathsf S$. We inserted a factor $1/8\pi G_N$ with $G_N$ the $D$-dimensional Newton constant, with $[G_N]=[\text{length}]^{{ D-2}}$, as an overall normalization convention.

\section{Perturbative expansion and quadratic spectra}\label{sec:perturbative}

	We now consider the perturbative expansion of the general EFT \eqref{eq:generalEFT} around a background. To that end, we solve the constraints \eqref{eq:invarconstraint} and parametrize the jets \eqref{eq:fulljetdata} as:
	\begin{align}\label{eq:restrictedEdgeDiffeomorphism}
		\Phi_0^a(x)&=\phi^a(x)\;,\qquad
		\Phi_0^i(x)=y^i(x)\coloneqq\left(e^{A^j(x)\partial_j}x\right)^i\;, \qquad J_a{}^\mu(x) = \delta_a^{\mathsf a}m_{\mathsf b}^{\mu}[\Phi_0]\Lambda^{\mathsf b}{}_{\mathsf a}\left[ \chi(x)\right]\;.
	\end{align}
	Here we have the $SO(1,1)$ boost matrix
	\begin{align}\label{eq:normalboostmatrix}
		\Lambda^{\mathsf a}{}_{\mathsf b}\left[ \chi\right] =
		\begin{pmatrix}
			\cosh\chi&-\sinh\chi\\
			-\sinh\chi&\cosh\chi
		\end{pmatrix}\;,
	\end{align}
	and $m_{\mathsf b}^{\mu}[\Phi_0]$ is an adapted orthonormal normal frame of the embedding $\Phi_0$ constructed as follows. First, we project the coordinate vectors $\delta_{\mathsf a}^{a}\partial_{\sigma^a}|_{\Phi_0}$ onto the normal plane, 
	\begin{align}\label{eq:projectedCoordinateNormals}
		\widetilde m_{\mathsf a}^{\mu}
		&=\delta_{\mathsf a}^{a}\left[
		\delta_a^{\mu}
		-\left(\eta_{ab}+\frac{\Phi_{0a}\Phi_{0b}}{\ell^2-\Phi_0^2}\right)
		\partial_i\Phi_0^b\,\left[(h^{(\Phi)})^{-1}\right]^{ij}\partial_j\Phi_0^{\mu}
		\right]\;.
	\end{align}
	 The second term subtracts the tangential component, so that $g_0(\widetilde m_{\mathsf a},t_i^{(\Phi)})=0$. We then normalize this and obtain 
	\begin{align}\label{eq:adaptedReferenceFrame}
		m_{\mathsf a}
		&=\widetilde m_{\mathsf b}\left[(\eta^{-1}\mathcal G)^{-1/2}\right]^{\mathsf b}{}_{\mathsf a}\;,\qquad
		\mathcal G_{\mathsf a\mathsf b}=g_0(\widetilde m_{\mathsf a},\widetilde m_{\mathsf b})\;.
	\end{align}
	The square-root branch is fixed by $m_{\mathsf a}[\iota_0]=n_{\mathsf a}^{(0)}$. 
	
	In the parametrization \eqref{eq:restrictedEdgeDiffeomorphism}, the background configuration corresponds to $A_i=\phi^a =\chi=0$. In small-field expansion, \eqref{eq:restrictedEdgeDiffeomorphism} reproduces \eqref{eq:edgemodejetiden}. The diffeomorphism jets and thus the fluctuations $A_i,\phi^a,\chi$ transform according to \eqref{eq:fulljetfiniteaction}. For example, under an external $SO(1,1)$ boost generated by $L_{0,{ D}}$ with constant rapidity $\lambda$, we have
\begin{align}\label{eq:adaptedJetFiniteBoost}
	A'_i(x)=A_i(x)\;,\qquad \phi'^a(x)=\Lambda^a{}_b[\lambda]\phi^b(x)\;,\qquad \chi'(x)=\chi(x)+\lambda\;.
\end{align}
Similarly, under an external global rotation $R\in SO({ D-1})$, $A_i$ transforms nonlinearly as
\begin{align}\label{eq:adaptedJetExternalRotation}
	A'_i
	=A_i+v_i+\frac12\left(A^j{ \nabla_j}v_i-v^j{ \nabla_j}A_i\right)+O(vA^2,v^2)\;,
\end{align}
while $\phi^a$ and $\chi$ are invariant. Here $v$ is the infinitesimal Killing generator of $R$. {Finally,} under an unbroken diagonal rotation $\left( R,\bar R\right) \in SO({ D-1})_\text{diag}$, with $\bar x=\bar R^{-1}x$, $A_i$ transforms tensorially as a vector,\footnote{Had we parametrized the tangential 0-jets in other ways such as $y^i=x^i+A^i$, the fluctuation $A^i$ generally does not transform tensorially as a vector.} while $\phi^a$ and $\chi$ transform as scalars{,}
\begin{align}\label{eq:adaptedJetDiagonalRotation}
	A'_i(x)=\frac{\partial\bar x^j}{\partial x^i}A_j(\bar x) \;, \qquad \phi'^a(x)=\phi^a(\bar x)\;,\qquad \chi'(x)=\chi(\bar x) \;.
\end{align}

In the parametrization \eqref{eq:restrictedEdgeDiffeomorphism}, the Biot strain \eqref{eq:straintensor} is independent of $\chi$, while the $\chi$-dependence in extrinsic curvature  \eqref{eq:extrinsiccurvature} and normal-bundle connection \eqref{eq:normalbundleconnection} cleanly separates out,
		\begin{gather}
			\mathsf S^i{}_j \left[ A,\phi,\chi\right] = \mathsf S^i{}_j \left[ A,\phi\right] \;, \qquad  K^{\mathsf a}{}_{ij}\left[ A,\phi,\chi\right] =\bigl(\Lambda^{-1}(\chi)\bigr)^{\mathsf a}{}_{\mathsf b}K^{\mathsf b}{}_{ij}\left[ A,\phi,0\right]  \;,  \nn\\
			\mathcal A_i\left[ A,\phi,\chi\right] =\mathcal A_i\left[ A,\phi,0\right] -{ \nabla_i}\chi \;. 
	\end{gather}
	
	\subsection{Small amplitude expansion and quadratic spectra}

	We now go back to our original motivation, namely to find a nonlinear completion of the theory that gives \eqref{eq:Zedgegravresult}. More specifically, we are looking for a fully nonlinear action $\mathcal{I}[A,\phi,\chi]$ such that
	\begin{align}
		Z_{\rm edge }&=\frac{1}{\mathcal{W}_\text{edge}\big|_\text{1-loop}}\;, \qquad \mathcal{W}_\text{edge}=\int \mathcal{D}A\,\mathcal{D}\phi \,\mathcal{D}\chi\, e^{-\mathcal{I}[A,\phi,\chi]} \;. 
	\end{align}
	
	Based purely on symmetries, we have arrived at the class of nonlinear theories \eqref{eq:geometricGradientAction}. To proceed further, we parametrize the typical field amplitudes by $|A_i|, |\phi^a| \sim \epsilon\ell $ and $|\chi|\sim \epsilon$. For the small-gradient expansion \eqref{eq:geometricGradientAction}, we keep $\epsilon$  fixed as $\ell$ grows.
	
	Now, we further take $\epsilon\ll1$ and expand the nonlinear action to quadratic order. At the lowest order in the small-gradient expansion, {i.e.,} \eqref{eq:leadingStrainAction}, only the following terms contribute through quadratic order in $\epsilon$:
	\begin{align}\label{eq:strainPotentialExpansion}
		B(\mathsf S)=c_0+c_1\tr\mathsf S
		+c_2(\tr\mathsf S)^2+c_3\tr(\mathsf S^2)+O(\mathsf S^3)\;.
	\end{align}
	On the other hand, the kinetic term for $\chi$ first appears at $N=2$ in the expansion \eqref{eq:geometricGradientAction}:
		\begin{align}\label{eq:rapidityKineticAction}
			I^{(2)}&\supset\frac{\ell^2 b_\chi}{8\pi G_N}\int_{S^{D-2}}\,
			\mathcal A_i[A,\phi,\chi]\mathcal A^i[A,\phi,\chi]\nn\\
			&=\frac{\ell^2 b_\chi}{8\pi G_N}\int_{S^{D-2}}\,
			\left[ \nabla_i\chi\nabla^i\chi-\epsilon^{(\perp)}_{ab}\nabla^i\chi\nabla^k\phi^a
			\left(\nabla_i\nabla_k\phi^b+\frac{\gamma_{ik}}{\ell^2}\phi^b\right)+O(\epsilon^4/\ell^2)\right] \;{,}
		\end{align}
	{where} $b_\chi$ is a dimensionless coupling.

Let us look at the kinetic terms for $A_i$ and $\phi^a$ in more detail. Here $\mathcal L_A$ denotes the Lie derivative along $A=A^i\partial_i$. Substituting the expansion
	\begin{align}\label{eq:dS4Sblocks}
			\mathsf S_{ij} \left[ A,\phi\right] 
			&={ \nabla_{(i}}A_{j)}
			+\frac14(\mathcal L_A^2\gamma)_{ij}
			-\frac18
			(\mathcal L_A\gamma)_{ik}
			(\mathcal L_A\gamma)_j{}^k+\frac12\eta_{ab} \left( { \nabla_i}\phi^{a}{ \nabla_j}\phi^{b}
			-\frac{\gamma_{ij}}{\ell^2}\phi^{a}\phi^{b}\right)\nn\\
			&\qquad+\frac1{12}(\mathcal L_A^3\gamma)_{ij}
			-\frac18(\mathcal L_A\gamma)_{k(i}(\mathcal L_A^2\gamma)_{j)}{}^k+\frac1{16}(\mathcal L_A\gamma)_i{}^k
			(\mathcal L_A\gamma)_k{}^l(\mathcal L_A\gamma)_{lj}\nn\\
			&\qquad-\frac14\eta_{ab}\left[
			(\mathcal L_A\gamma)_{k(i}{ \nabla_{j)}}\phi^a{ \nabla^k}\phi^b
			+\frac{\phi^a\phi^b}{\ell^2}(\mathcal L_A\gamma)_{ij}
			\right]+O(\epsilon^4)\;.
		\end{align}
		 into \eqref{eq:strainPotentialExpansion} and discarding a total derivative term gives
	\begin{align}\label{eq:dS4expandedaction}
		I^{(0)}
		&=\frac{1}{8\pi G_N}\int_{S^{{ D-2}}}\,
		\Bigg[c_0+
		\left(\frac{c_1}{2}+c_3\right)
		\left(\frac14F_{ij}F^{ij}-{ \frac{{ D-3}}{\ell^2}}A_iA^i\right)
		+(c_2+c_3)({ \nabla_i}A^i)^2\nn\\
		&\qquad+\frac{c_1}{2}\eta_{ab} 
		\left({ \nabla_i}\phi^{a}{ \nabla^i}\phi^b
		-{ \frac{{ D-2}}{\ell^2}}\phi^a\phi^b\right)+O(\epsilon^3)
		\Bigg]\;{,}
	\end{align}
	where $F_{ij}=2{ \nabla_{[i}}A_{j]}$. For generic $c_2$ and $c_3$, the longitudinal part of the vector, {i.e.,} $A^L_i=\partial_i \psi$ with scalar $\psi$, has an independent kinetic term through $({ \nabla_i}A^i)^2$. Therefore, for the theory to reproduce the spectra \eqref{eq:Zedgegravresult}, we must set $c_3=-c_2$. 
	
	Thus, a large class of actions $\mathcal{I}[A,\phi,\chi]$ that reproduce the quadratic operators in \eqref{eq:Zedgegravresult} at the lowest nonvanishing order in the small-gradient expansion for each field is given by  \eqref{eq:geometricGradientAction} with $c_3=-c_2$ and non-zero $c_1$, $c_1/2-c_2$ and $b_\chi$.\footnote{As pointed out in \cite{Law:2025ktz}, the relative sign of the timelike and spacelike bending modes $\phi^0$ and $\phi^D$ might explain the overall phase $i^{D+2}$ of the full one-loop sphere gravitational path integral \cite{Polchinski:1988ua}.}  These EFTs are interacting multi-field theories of shift-symmetric fields on $S^{D-2}$, containing two $k=1$ scalars $\phi^a$, one $k=0$ scalar $\chi$, and
	one $k=0$ vector $A_i$, where $k$ denotes the shift level. They generalize
	the single-field scalar EFTs studied in
	\cite{Bonifacio:2018zex,Bonifacio:2021mrf} and the single-field spin-1 EFT of \cite{Bonifacio:2019hrj}.
			
	 The tuning $c_3=-c_2$ combines the quadratic Biot strain terms in \eqref{eq:strainPotentialExpansion} into a term proportional to the second elementary symmetric polynomial $e_2(\mathsf S)=\frac12\left[(\operatorname{tr}\mathsf S)^2	-\operatorname{tr}(\mathsf S^2)\right]$. This suggests a natural nonlinear generalization:
	\begin{align}\label{eq:strainPotentialcontrain}
		B(\mathsf S)=\sum_{n=0}^{D-2} \beta_n e_n\left(\mathsf S \right)\;{,}
	\end{align}
	where $e_n$ are elementary symmetric polynomials.\footnote{Elementary symmetric polynomials for a $d$-by-$d$ matrix $\mathsf S$ are defined by $\det(I+t\mathsf S)=\sum_{n=0}^{d}t^n e_n(\mathsf S)$. Thus, we have $e_0(\mathsf S)=1$, $e_1(\mathsf S)=\operatorname{tr}\mathsf S$, $e_2(\mathsf S)=\frac12\left[(\operatorname{tr}\mathsf S)^2	-\operatorname{tr}(\mathsf S^2)\right]$, $e_3(\mathsf S)=\frac16\left[(\operatorname{tr}\mathsf S)^3
	-3(\operatorname{tr}\mathsf S)\operatorname{tr}(\mathsf S^2)
	+2\operatorname{tr}(\mathsf S^3)\right]${,} and so on. In particular, $e_d(\mathsf S)=\det\mathsf S$ and $e_n=0$ for $n>d$.} In any $D$, this consists of a finite number of terms. In particular, for $D=4$, $B(\mathsf S)=\beta_0+\beta_1\operatorname{tr}\mathsf S
	+\beta_2\det\mathsf S$. Curiously, this polynomial structure also appears in dRGT massive gravity \cite{deRham:2010kj,Hassan:2011hr} and related brane-like theories \cite{Gabadadze:2012tr,Andrews:2013ora,deRham:2015ijs}. We leave the investigation of the relation between the present edge mode theory and these theories to future work. Whether the restricted potential \eqref{eq:strainPotentialcontrain} is realized, and the values of its coefficients, must ultimately be determined by the underlying gravitational theory.

\section{Outlook}

We have identified gravitational edge modes \eqref{eq:shiftsymmetry} as Goldstones associated with the symmetry breaking \eqref{eq:edgesymbreak} induced by a framed horizon-sphere embedding in a de Sitter space. Based on this geometric picture, we constructed their effective field theories. We conclude by highlighting several interesting prospects.
	
	First of all, despite the geometric interpretation for the fields underlying \eqref{eq:Zedgegravresult} established in the present work, a first-principles derivation of the bulk-edge split \eqref{eq:Z1loopbulkedge} remains open. Extending the phase-space analysis in {Appendix}~\ref{app:CPS} and quantizing the resulting edge theory might identify the quantum quantity represented by $Z_{\rm edge}$. Presumably a top-down derivation starting from a given gravitational theory would determine the coefficients of the EFT \eqref{eq:generalEFT} of the edge mode sector and its (possibly non-local) couplings to bulk graviton fluctuations.
	
	Next, the intrinsic geometric notion of ``observer" stressed in this work aligns in spirit with existing work on dynamical reference frames \cite{Carrozza:2021gju,Carrozza:2022xut,Aguilar-Gutierrez:2026svf},\footnote{For example, the reference copy of $dS_{{ D}}$, equipped with the pullback metric $\Phi^*g_0$, plays a role analogous to the ``relational spacetime'' of \cite{Carrozza:2022xut}.} and it	would be interesting to establish the precise connection. Another important direction is to investigate the interplay of the gravitational edge modes with a physical observer, whose role in de Sitter quantum gravity is emphasized in recent discussions \cite{Chandrasekaran:2022cip,Maldacena:2024spf,Chen:2025jqm,Blommaert:2026ofx}. Initial steps in this direction appear in \cite{Chandrasekaran:2026pnc,Klinger:2026tws}.
	
	Lastly, the breaking of spacetime and internal symmetries to a diagonal subgroup in \eqref{eq:edgesymbreak} is familiar from condensed matter systems \cite{Nicolis:2013lma,Nicolis:2015sra,Esposito:2026diu}. This suggests a systematic EFT approach to cosmological horizon membranes \cite{Anninos:2011zn,Anninos:2018HolographicLiquids,Anninos:2024wpy,Anninos:2025zgr}, building on the membrane paradigm \cite{Price:1986yy,Parikh:1997ma,Susskind:1993if}, fluid-gravity correspondence \cite{Bhattacharyya:2007vjd,Bredberg:2011jq}, and earlier symmetry-breaking perspectives \cite{Eling:2016xlx}. Studying their macroscopic dynamics might guide us towards microscopic models of the de Sitter horizon~\cite{Banks:2003cg,Banks:2006rx,Banks:2013qpa}.

		\section*{Acknowledgments} 
		
		It is a great pleasure to thank Luca Ciambelli, Austin Joyce, Michael Landry, Hong Liu, Rachel Rosen and Douglas Stanford for stimulating conversations, and especially Dionysios Anninos, Laurent Freidel and Philipp H\"{o}hn for useful discussions and comments on the draft. AL was supported in part by the Stanford Science Fellowship, NSF grant PHY-2310429, U.S. Department of Energy grant DE-SC0026143, and Simons Foundation grants (926198, DS, 507021 ES). VL was supported in part by the U.S. Department of Energy under Grant No. DE-SC0011941. 
		
		 We acknowledge OpenAI's Codex for assistance with some initial calculations, literature search, and manuscript editing.
		 
		\appendix

			\section{Covariant phase space analysis}\label{app:CPS}
			
			{In this appendix, we perform} a preliminary covariant phase space analysis to address the physicality of the 0- and 1-jets \eqref{eq:edgemodejetiden} in Einstein gravity. To begin, we pick a Cauchy slice $\Sigma$ on the static patch background \eqref{eq:tubularcoor} with metric $g_0$ that ends on the bifurcation sphere $\mathcal B=\partial\Sigma{ =\iota_0(S^{{ D-2}}_{\rm int})}$. 
			
			We consider nearby on-shell metrics $g$, while keeping the slice embedding and its boundary restriction $\iota_0$ fixed. On the space of such metrics, the presymplectic potential current and presymplectic form are respectively \cite{Iyer:1994ys}\footnote{$\theta^\mu(g;\delta g)$ is defined only up to the divergence of a local antisymmetric tensor, which shifts $\Omega_\Sigma$ by a corner term on $\mathcal B$. We ignore this ambiguity below and leave its systematic treatment to future work; {see, e.g.,} \cite{Harlow:2019yfa,Freidel:2020xyx} for discussions.}
			\begin{align}\label{eq:EHpresymplectic}
				\theta^\mu(g;\delta g)
				&=\frac{1}{16\pi G_N}\left[
				{\nabla^{(g)}_\nu}\left(g^{\mu\rho}g^{\nu\sigma}\delta g_{\rho\sigma}\right)
				-{\nabla^{(g)\mu}}\left(g^{\rho\sigma}\delta g_{\rho\sigma}\right)
				\right]\;,\nn\\
				{\Omega_\Sigma(\delta_1g,\delta_2g)}
				&{ {}=\int_\Sigma\left[
				\delta_1\!\left(d\Sigma_\mu\,\theta^\mu(g;\delta_2g)\right)
				-\delta_2\!\left(d\Sigma_\mu\,\theta^\mu(g;\delta_1g)\right)
				\right]}\;.
			\end{align}
			In these expressions, $\nabla^{(g)}$ denotes the {Levi-Civita} connection of the varying spacetime metric $g$ {and} $\delta$ denotes the exterior derivative on field space, which acts on the metric $g$ but not on the slice embedding or $\iota_0$. In the presymplectic form, $d\Sigma_\mu\coloneqq -n^\Sigma_\mu\,dV_\Sigma$, where $n^\Sigma_\mu$ is the future-directed unit normal to $\Sigma$, and $dV_\Sigma$ is its induced volume element. Both depend on $g$ and are thus  acted upon by $\delta$ non-trivially.
			
			Now, consider the direction $\delta g=\mathcal L_\zeta g_{\mu\nu}$, {i.e.,} a diffeomorphism generated by {the} vector field $\zeta^\mu$, with $\delta\zeta^\mu=0$. 
			Plugging this into the presymplectic form leads to a corner term at $\mathcal B=\partial\Sigma$ on shell:
			\begin{align}\label{eq:cornerpairing}
				\Omega_\Sigma(\delta g,\mathcal L_\zeta g)\big|_{\rm EOM}
				&=-\frac{1}{8\pi G_N}
				\delta\!\int_{\mathcal B}d^{{ D-2}}x\,\sqrt{h}\,
				n_{0[\mu}n_{{ D},\nu]}{\nabla^{(g)\mu}}\zeta^\nu\nn\\
				& \qquad +2\int_{\mathcal B}d^{{ D-2}}x\,\sqrt{h}\,
				n_{0[\mu}n_{{ D},\nu]}\zeta^\mu\theta^\nu(g;\delta g)\;{.}
			\end{align}
			Here $n_0,n_{{ D}}$ are orthonormal with respect to $g$, with $n_{0\mu}=n^\Sigma_\mu|_{\mathcal B}$ and $n_{{ D}}$ the inward unit normal to $\mathcal B$ within $\Sigma$. We choose $\Sigma$ so that $n_{\mathsf a}=n_{\mathsf a}^{(0)}$ at $g=g_0$, with $n_{\mathsf a}^{(0)}$ as in \eqref{eq:normalframe}. $h_{ij}$ is the metric induced on $\mathcal B$ and $h$ its determinant. 
			
			From \eqref{eq:cornerpairing}, we see that a ``small'' diffeomorphism with zero 0- and 1-jets corresponds to a degenerate direction.  The mixed first jets $\partial_a\zeta^i|_{\mathcal B}$
				 also drop out of \eqref{eq:cornerpairing} and thus are null directions of the presymplectic form \eqref{eq:EHpresymplectic}. In contrast, a ``large" diffeomorphism with nonzero 0- or 1-jets at $\mathcal B$ need not be a degenerate direction of the subregion presymplectic form. 
			
			Explicitly, evaluating \eqref{eq:cornerpairing} on the background \eqref{eq:tubularcoor} with the specific 0- and 1-jets \eqref{eq:edgemodejetiden} gives
			\begin{align}\label{eq:cornerpairinglowjets}
				\Omega_\Sigma(\delta g,\mathcal L_\zeta g)\big|_{{\rm EOM},g=g_0}
				&{{}=\frac{1}{16\pi G_N}\int_{\mathcal B}d^{{ D-2}}x\,
				\left[
				2\chi\,\delta\sqrt{h}+A^i\delta\mathcal F_i
				-2\sqrt{\gamma}\,\phi^a\epsilon^{(\perp)}_{ab}\delta K^b
				\right]_{g=g_0}}\;,
			\end{align}
			where 
			\begin{align}\label{eq:cornernormaldata}
				{\mathcal F_i\coloneqq\left.\sqrt{h}\,g_{i\mu}[n_0,n_{{ D}}]^\mu\right|_{\mathcal B}
				\;,\qquad K^a\coloneqq-\left.h^{ij}n_{\mathsf b}^{\,a}\,g_{i\mu}{\nabla^{(g)}_j} n^{\mathsf b\mu}\right|_{\mathcal B}}\;.
			\end{align}
			Here $K^a$ are the transverse components of the mean-curvature vector. %
				For the ``twist density vector" $\mathcal F_i$ \cite{Donnelly:2016auv}, we hold the {coordinates} \eqref{eq:tubularcoor} fixed as $g$ varies and extend the normals as $g$-orthonormal normals to the constant-$\sigma^a$ surfaces. In \eqref{eq:cornerpairinglowjets}, the variations are taken before setting $g=g_0$.

At linear order about $g_0$, \eqref{eq:cornerpairinglowjets} lets us read off the conjugate pairings
			\begin{align}
				\chi \leftrightarrow \frac{\sqrt{h}}{8\pi G_N} \;, \qquad
				A^i \leftrightarrow \frac{\mathcal F_i}{16\pi G_N} \;, \qquad
				{\phi^a \leftrightarrow -\frac{\sqrt{\gamma}}{8\pi G_N}\epsilon^{(\perp)}_{ab}K^b}\;,
			\end{align}
			corresponding to the area density, the twist density vector, and the dual mean-curvature density, respectively. We stress that these pairings concern the linearized fluctuations of these quantities.  At the full nonlinear level, it is known that the charge variations associated with the normal 0-jets $\phi^a$ are generally non-integrable; see \cite{Speranza:2017gxd,Freidel:2021cjp} for discussions.

			While this discussion shows that the low jets \eqref{eq:edgemodejetiden} {\it could} be physical, whether they are retained as genuine phase-space variables depends on the choice of boundary condition at $\mathcal B=\partial\Sigma$ and on a prescription for treating the corner ambiguities in \eqref{eq:EHpresymplectic}. A related comment is that among all the possibly physical 1-jets, {i.e.,} the traceless part of the matrix $\partial_a \xi^b|_{\sigma=0}$, which parametrize an element of $\mathfrak{sl}\left( 2,\mathbb{R}\right) $ at any given point on the corner, only the $\mathfrak{so}(1,1)$ part is retained. A possible perspective is that part of the boundary condition fixes $g(n_{\mathsf a}^{(0)},n_{\mathsf b}^{(0)})|_{\mathcal B}=\eta_{\mathsf a\mathsf b}$, with the reference normals $n_{\mathsf a}^{(0)}$ held fixed. At $g=g_0$, preserving this condition restricts the normal 1-jets to $\mathfrak{so}(1,1)$. We leave these issues, together with the quantization of the resulting phase space, to future work. The gauge-theory analyses in \cite{Ball:2024gti,Ball:2024hqe,Ball:2024xhf} could serve as inspiration.

			\section{Edge mode theory from the coset construction}\label{sec:coset}
		
		In this appendix, we systematically construct effective field theories (EFTs) for the symmetry breaking pattern \eqref{eq:edgesymbreak} using the so-called coset construction, developed first for broken internal symmetries \cite{Coleman:1969sm,Callan:1969sn} and later for spacetime symmetries \cite{Volkov:1973vd,Ogievetsky1974}. This systematic method takes the symmetry-breaking pattern as input and organizes the invariant terms in a derivative expansion. Modern applications to condensed-matter systems \cite{Nicolis:2013lma,Nicolis:2015sra} and brane-like theories \cite{Clark:2005ht,Clark:2006pd,Clark:2007rn,Delacretaz:2014oxa} are particularly close in spirit.

		We begin with a standard basis ${L_{\mathsf A\mathsf B}=-L_{\mathsf B\mathsf A}}$  with ${\mathsf A,\mathsf B=0,1,\dots,D}$ for $\mathfrak{so}(1,D)_\text{ext}$, which obeys
		\begin{align}\label{eq:SO14comm}
			{\left[ L_{\mathsf A\mathsf B},L_{\mathsf C\mathsf D}\right] = \eta_{\mathsf B\mathsf C}L_{\mathsf A\mathsf D}-\eta_{\mathsf A\mathsf C}L_{\mathsf B\mathsf D}+\eta_{\mathsf A\mathsf D}L_{\mathsf B\mathsf C}-\eta_{\mathsf B\mathsf D}L_{\mathsf A\mathsf C}} \;, 
		\end{align}
		and similarly  ${\bar L_{\mathsf I\mathsf J}=-\bar L_{\mathsf J\mathsf I}}$  with ${\mathsf I,\mathsf J=1,\dots,D-1}$  for $ \mathfrak{so}(D-1)_\text{int}${,}
		\begin{align}
			{\left[ \bar L_{\mathsf I\mathsf J},\bar L_{\mathsf K\mathsf L}\right] = \delta_{\mathsf J\mathsf K}\bar L_{\mathsf I\mathsf L}-\delta_{\mathsf I\mathsf K}\bar L_{\mathsf J\mathsf L}+\delta_{\mathsf I\mathsf L}\bar L_{\mathsf J\mathsf K}-\delta_{\mathsf J\mathsf L}\bar L_{\mathsf I\mathsf K}} \;{.}
		\end{align}	
		
			\paragraph{Unbroken generators}
		
		To proceed, we {rename} the following generators{:} 
		\begin{gather}\label{eq:dS4MABdictionary}
			J_{\mathsf{ij}} =L_{\mathsf{ij}} \; ,\quad
			{ S_{\mathsf i}=\frac1\ell L_{{\mathsf i},D-1}} \; ,\quad H=L_{0,D} \;, \quad 
			{ Z_{\mathsf a}=\frac1\ell L_{\mathsf a,D-1}}\;,\quad
			 \bar J_{\mathsf{ij}} =\bar L_{\mathsf{ij}} \; ,\quad
			{ \bar S_{\mathsf i}=\frac1\ell\bar L_{{\mathsf i},D-1}}\; ,
		\end{gather}
		where ${\mathsf i}=1,\dots,D-2$ and $\mathsf a=0,D$.
		The unbroken subgroup $SO(D-1)_\text{diag}$ is generated by
		\begin{align}\label{eq:edgediagen}
			T_{{\mathsf i}} = S_{\mathsf i} +\bar S_{\mathsf i} \;, \qquad U_{\mathsf{ij}} = J_{\mathsf{ij}} +\bar J_{\mathsf{ij}}  \;. 
		\end{align}
			We view $T_{\mathsf i}$ as the unbroken translations along the diagonal sphere $S^{D-2}_\text{diag}\coloneqq \{ \left(p,\iota(p) \right) : p \in S^{D-2}_\text{int} \}\subset S^{D-2}_\text{int} \times dS_D$ in the background configuration. After choosing an origin $\left({p_o},\iota({p_o}) \right) $ on $S^{D-2}_\text{diag}$, the linearly realized local subgroup is the $SO(D-2)_\text{diag}$ generated by $U_{\mathsf{ij}}$. The commutators with the broken generators just listed, together with $L_{\mathsf a \mathsf i}$, are
		\begin{gather}
			{ [T_{\mathsf i},S_{\mathsf j}]=-\frac1{\ell^2}J_{\mathsf i\mathsf j}} \; ,\quad
			[T_{\mathsf i},J_{\mathsf j \mathsf k}]=\delta_{\mathsf i\mathsf j}S_{\mathsf k}-\delta_{\mathsf i\mathsf k}S_{\mathsf j}	\; , \quad
			[T_{\mathsf i},H]=0\;, \nn\\
			{ [T_{\mathsf i},Z_{\mathsf a}]=\frac1{\ell^2}L_{\mathsf a\mathsf i}}\; ,\quad
			[T_{\mathsf i},L_{\mathsf a{\mathsf j}}]=-\delta_{\mathsf i \mathsf j}Z_{\mathsf a}\; .\label{eq:dS4mixedcomm}
		\end{gather}

		\paragraph{Local section}
		
		Proceeding as in the standard coset construction for spacetime symmetries, we group the sphere coordinates and Goldstone fields together in an enlarged coset
\begin{align}\label{eq:edgequotientspace}
V(x)\in \frac{SO(1,D)_\text{ext}\times SO(D-1)_\text{int}}{SO(D-2)_\text{diag}}
\stackrel{\text{locally}}{\simeq}
S^{D-2}_\text{diag}\times
\frac{SO(1,D)_\text{ext}\times SO(D-1)_\text{int}}{SO(D-1)_\text{diag}}\; ,
\end{align}
so that the first factor contains the coordinates while the Goldstones take values in the second factor.
		Viewing $SO(1,D)_\text{ext}\times SO(D-1)_\text{int}$ as a principal $SO(D-2)_\text{diag}$ bundle over \eqref{eq:edgequotientspace}, we choose a local representative
		\begin{align}\label{eq:localrepresentative}
			V(x)=\left(g(x),s(x)\right)\;,\qquad
			g(x)\in SO(1,D)_\text{ext}\;,\quad s(x)\in SO(D-1)_\text{int}\;. 
		\end{align}
		For example, assigning $A^i,\chi,\phi^{\mathsf a},\beta^{{\mathsf a}{\mathsf i}}${, and} $\alpha^{\mathsf i\mathsf j}=-\alpha^{\mathsf j\mathsf i}$ to be the Goldstones associated with $S_{\mathsf i},H,\allowbreak Z_{\mathsf a},L_{\mathsf a{\mathsf i}}${, and} $J_{\mathsf i\mathsf j}${,} respectively, we might choose the (non-unique) explicit parametrization
\begin{align}\label{eq:dS4cosetrepmain}
V
&=\left({ e^{q^{\mathsf i}(y)S_{\mathsf i}}}
e^{-\chi H}{ e^{\phi^{\mathsf a}Z_{\mathsf a}}}
e^{\beta^{\mathsf a\mathsf i}L_{\mathsf a\mathsf i}}
e^{\frac12\alpha^{\mathsf i\mathsf j}J_{\mathsf i\mathsf j}},\;
{ e^{q^{\mathsf i}(x)\bar S_{\mathsf i}}}\right)\;,
\end{align}
where $y^i\coloneqq x^i+A^i(x)$\footnote{This differs from the flow parametrization \eqref{eq:restrictedEdgeDiffeomorphism} by a field redefinition starting at quadratic order.} and $q^{\mathsf i}$ are exponential coordinates on $S^{D-2}$ so that $q^{\mathsf i}(x({p_o}))=0$. As a representative of an element of \eqref{eq:edgequotientspace}, it is subject to the equivalence relation under the right action
		\begin{align}\label{eq:U(1)}
			V(x)\sim V(x) h^{-1}(x) \;, \qquad h(x) \in SO(D-2)_\text{diag} \;. 
		\end{align}
		For a given parametrization such as \eqref{eq:dS4cosetrepmain}, the global symmetry \eqref{eq:fullsymmetry} is nonlinearly realized as 
\begin{align}\label{eq:globalsym}
V\mapsto G\cdot V\cdot h_G^{-1}(x)\;,
\end{align}
where $G\in SO(1,D)_\text{ext}\times SO(D-1)_\text{int}$ acts on the left and $h_G(x)=e^{\frac12\lambda_G^{\mathsf i\mathsf j}(x)U_{\mathsf i\mathsf j}} \in SO(D-2)_\text{diag}$ is a compensating right rotation that restores the chosen local representative \eqref{eq:dS4cosetrepmain}. With \eqref{eq:globalsym}, one can work out the field transformations for any given parametrization, at least order by order. For \eqref{eq:dS4cosetrepmain}, these reduce at lowest order to field-independent shifts, reproducing \eqref{eq:jettransform} for $A^i,\phi^{\mathsf a}${, and} $\chi$.

		\subsection{Maurer-Cartan form}
		
		The fundamental object of the coset construction is the Maurer-Cartan (MC) form (or more precisely its pullback to the intrinsic round $S^{ D-2}_\text{int}$).
		\begin{align}\label{eq:MCdef}
			\Omega \coloneqq V^{-1}dV \;. 
		\end{align}
		Under the nonlinearly realized global symmetry \eqref{eq:globalsym}, the constant left action drops out of $V^{-1}dV$; only the compensating right $SO(D-2)_\text{diag}$ rotation acts on the MC form:
		\begin{align}\label{eq:MC transform}
			\Omega \mapsto { h_G} \,\Omega \,{ h_G^{-1}} +{ h_G}\,d{ h_G^{-1}} \;. 
		\end{align}
		As a 1-form valued in the Lie algebra $\mathfrak{so}(1,D)_\text{ext}\oplus \mathfrak{so}( D-1)_\text{int}$, \eqref{eq:MCdef} can be expanded in their basis:\footnote{For an explicit representative such as \eqref{eq:dS4cosetrepmain}, the MC coefficients follow by repeatedly applying the identities %
			$e^{-X}de^X=\int_0^1du\,e^{-uX}(dX)e^{uX}
			=\frac{1-e^{-\mathrm{ad}_X}}{\mathrm{ad}_X}(dX)$ and $e^{-X}Ye^X=e^{-\mathrm{ad}_X}Y$.}
\begin{align}\label{eq:MCexpand}
\Omega&=\bar\omega_S^{\mathsf i}\bar S_{\mathsf i}
+\frac12\bar\omega_J^{\mathsf i\mathsf j}\bar J_{\mathsf i\mathsf j}
+\omega_S^{\mathsf i}S_{\mathsf i}
+\frac12\omega_J^{\mathsf i\mathsf j}J_{\mathsf i\mathsf j}
+\omega_Z^{\mathsf a}Z_{\mathsf a}
+\omega^{\mathsf a\mathsf i}L_{\mathsf a\mathsf i}+\omega_HH\nn\\
&=\omega_T^{\mathsf i}T_{\mathsf i}
+\frac12\omega_U^{\mathsf i\mathsf j}U_{\mathsf i\mathsf j}
+\Delta\omega_S^{\mathsf i}S_{\mathsf i}
+\frac12\Delta\omega_J^{\mathsf i\mathsf j}J_{\mathsf i\mathsf j}
+\omega_Z^{\mathsf a}Z_{\mathsf a}
+\omega^{\mathsf a\mathsf i}L_{\mathsf a\mathsf i}+\omega_HH\;.
\end{align}
		Thus $\omega_T^{\mathsf i}={\bar\omega_S}^{\mathsf i}$, $\omega_U^{\mathsf i\mathsf j}=\bar\omega_J^{\mathsf i\mathsf j}$,
		$\Delta\omega_S^{\mathsf i}=\omega_S^{\mathsf i}-{\bar\omega_S}^{\mathsf i}$, and
		$\Delta\omega_J^{\mathsf i\mathsf j}=\omega_J^{\mathsf i\mathsf j}-\bar\omega_J^{\mathsf i\mathsf j}$.  
From \eqref{eq:MC transform}, the coefficients of the broken generators transform homogeneously under the compensating $SO(D-2)_\text{diag}$ rotation, 
	\begin{align}\label{eq:MCcomponentTransformations}
	\Delta\omega_S^{\mathsf i}\mapsto (h_G)^{\mathsf i}{}_{\mathsf j}\Delta\omega_S^{\mathsf j}\;,\quad 
		\Delta\omega_J\mapsto h_G\Delta\omega_J h_G^{-1}\;,\quad
		\omega^{\mathsf a\mathsf i}\mapsto (h_G)^{\mathsf i}{}_{\mathsf j}\omega^{\mathsf a\mathsf j}\;,\quad 
		\omega_Z^{\mathsf a}\mapsto\omega_Z^{\mathsf a}\;,\quad\omega_H\mapsto\omega_H\;,
	\end{align}
	while the unbroken coefficients $\omega_T$ and $\omega_U$ transform as a vector and connection, respectively, %
\begin{align}\label{eq:MCcomponentTransformationsunbroken}
\omega_T^{\mathsf i}\mapsto (h_G)^{\mathsf i}{}_{\mathsf j}\omega_T^{\mathsf j}\;, \qquad 
\omega_U\mapsto h_G\omega_U h_G^{-1}+h_G\,dh_G^{-1}\;.
\end{align}
In fact, the unbroken coefficients give the round coframe and spin connection on $S^{D-2}_\text{diag}$:
\begin{align}\label{eq:veilbeinconnection}
{ \omega_T^{\mathsf i}=\bar\Theta^{\mathsf i}(x)}\;,
\qquad\omega_U^{\mathsf i\mathsf j}=\omega^{\mathsf i\mathsf j}(x)\;.
\end{align}
For the parametrization \eqref{eq:dS4cosetrepmain}, these are
\begin{align}\label{eq:intrinsicSphereForms}
\bar\Theta^{\mathsf i}
&=\left[\frac{\sin(|q|/\ell)}{|q|/\ell}\delta^{\mathsf i}{}_{\mathsf j}
+\left(1-\frac{\sin(|q|/\ell)}{|q|/\ell}\right)
\frac{q^{\mathsf i}q_{\mathsf j}}{|q|^2}\right]dq^{\mathsf j}\;, \qquad 
\omega^{\mathsf i\mathsf j}
=\frac{1-\cos(|q|/\ell)}{|q|^2}
\left(q^{\mathsf i}dq^{\mathsf j}-q^{\mathsf j}dq^{\mathsf i}\right)\;,
\end{align}
where $q^{\mathsf i}=q^{\mathsf i}(x)$ and $|q|^2=\delta_{\mathsf i\mathsf j}q^{\mathsf i}q^{\mathsf j}$, with the expressions understood by continuity at $q=0$.
		
		\subsection{Inverse Higgs constraints and their geometric interpretations}\label{sec:IHC}
		
		In systems with spontaneously broken spacetime symmetries, the number of broken generators can exceed the number of independent Goldstone fields required for the nonlinear realization \cite{Ivanov:1975zq,Low:2001bw,McArthur:2010zm,Nicolis:2013sga,
			Brauner:2014aha,Klein:2017npd}. The extra redundant Goldstones may be eliminated by the so-called inverse Higgs constraints (IHCs). In the present case, we focus on the constraints arising from the two commutators%
\begin{gather}\label{eq:IHCcomm}
[T_{\mathsf i},L_{\mathsf a\mathsf j}]=-\delta_{\mathsf i\mathsf j}Z_{\mathsf a}\;,
\qquad[T_{\mathsf i},J_{\mathsf j\mathsf k}]
=\delta_{\mathsf i\mathsf j}S_{\mathsf k}-\delta_{\mathsf i\mathsf k}S_{\mathsf j}\;.
\end{gather}
		The first one permits an IHC that eliminates the Goldstones associated with $L_{\mathsf a{\mathsf j}}$:
		\begin{align}\label{eq:IHC1}
			\omega_Z^{\mathsf a} =0 \;.
		\end{align}
		To state the constraint arising from the second commutator \eqref{eq:IHCcomm}, expand 
		\begin{align}\label{eq:omegaSexpand}
			{ \Delta\omega_S^{\mathsf i}=\left(M^{\mathsf i}{}_{\mathsf j}-\delta^{\mathsf i}{}_{\mathsf j}\right)\bar\Theta^{\mathsf j}(x)} \;.
		\end{align}
		The Goldstones associated with $J_{\mathsf i\mathsf j}$ could be eliminated by setting 
		\begin{align}\label{eq:IHC2}
			M_{[{\mathsf i \mathsf j}]}=0 \;.  
		\end{align}
	The constraints \eqref{eq:IHC1} and \eqref{eq:IHC2} are locally solvable algebraically: \eqref{eq:IHC1} determines  the $2(D-2)$ mixed-frame variables $\beta^{\mathsf a\mathsf i}$ in terms of $A^i,\phi^{\mathsf a},\chi$ and their first derivatives. The remaining tangent rotations $\alpha^{\mathsf i\mathsf j}$ are fixed by \eqref{eq:IHC2}. The IHCs admit natural geometric interpretations. Similar discussions in other contexts can be found in \cite{Delacretaz:2014oxa}.

		To understand the first IHC \eqref{eq:IHC1}, it is useful to use the ambient
		realization of $dS_{{ D}}$ 
\begin{align}\label{eq:dS4embeddingspace}
\eta_{AB}X^AX^B=-(X^0)^2+\sum_{A=1}^{D}(X^A)^2=\ell^2\;,
\qquad A,B=0,1,\ldots,D\;,
\end{align}
		and the matrix representation of the $SO(1,{ D})$ generators, 
\begin{align}
\left(\mathbf L_{\mathsf A\mathsf B}\right)^{\mathsf C}{}_{\mathsf D}
\coloneqq\delta^{\mathsf C}_{\mathsf A}\eta_{\mathsf B\mathsf D}
-\delta^{\mathsf C}_{\mathsf B}\eta_{\mathsf A\mathsf D}\;.
\end{align}
We choose an origin  $p_o\in S^{{ D-2}}_\text{int}$ with embedding space representative
\begin{align}\label{eq:referencepoint}
{ (\mathcal X_o)^A\coloneqq\ell\,\delta^A_{D-1}}\;.
\end{align}
		The generators ${\mathbf J_{\mathsf i\mathsf j}\coloneqq\mathbf L_{\mathsf i\mathsf j}}$, ${\mathbf H\coloneqq\mathbf L_{0,D}}$ and ${\mathbf L}_{\mathsf a{\mathsf i}}$ preserve this point
		while ${ \mathbf S_{\mathsf i}\coloneqq\mathbf L_{\mathsf i,D-1}/\ell}$ and ${ \mathbf Z_{\mathsf a}\coloneqq\mathbf L_{\mathsf a,D-1}/\ell}$
		\begin{align}\label{eq:referenceframe}
			{ { \left(\mathbf S_{\mathsf i}\mathcal X_o\right)^A=\delta^A_{\mathsf i}}\;,
\qquad { \left(\mathbf Z_{\mathsf a}\mathcal X_o\right)^A=\delta^A_{\mathsf a}}}\;,
		\end{align} 	
		define tangent and normal frames to the bifurcation sphere at $X^0=X^D=0$ at $\mathcal{X}_o$ respectively.
		These frames are orthonormal 
		\begin{align}
			  (\mathbf S_{\mathsf i}\mathcal X_o)\cdot(\mathbf S_{\mathsf j}\mathcal X_o)=\delta_{\mathsf i\mathsf j}\;,\qquad
 (\mathbf S_{\mathsf i}\mathcal X_o)\cdot(\mathbf Z_{\mathsf a}\mathcal X_o)=0\;,\qquad
(\mathbf Z_{\mathsf a}\mathcal X_o)\cdot(\mathbf Z_{\mathsf b}\mathcal X_o)=\eta_{\mathsf a\mathsf b}\;,
		\end{align}
		where $\cdot$ denotes the contraction with the ambient $(D+1)$-dimensional Minkowski metric \eqref{eq:dS4embeddingspace} and $\eta_{\mathsf a\mathsf b}=\text{diag}(-1,1)$ is the 2D Minkowski metric in the $(X^0,X^{{ D}})$-plane. 
		
\begin{figure}[H]
			\centering
			\includegraphics[width=\linewidth]{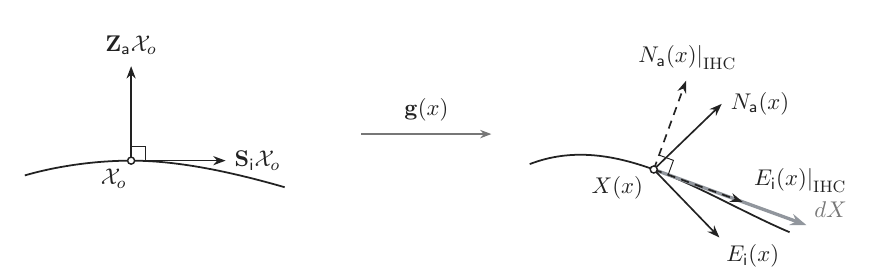}
			\caption{Under $\mathbf g(x)$, the orthonormal reference frame $\{\mathbf S_{\mathsf i}\mathcal X_o,\mathbf Z_{\mathsf a}\mathcal X_o\}$ at $\mathcal X_o$ is mapped to the orthonormal vectors $E_{\mathsf i}(x)$ and $N_{\mathsf a}(x)$ at $X(x)$, which need not be adapted to the deformed surface. The first IHC $\omega_Z^{\mathsf a}=0$ adapts this frame (dashed), making $E_{\mathsf i}$ tangent and $N_{\mathsf a}$ normal to the surface, so that $dX=\Theta^{\mathsf i}E_{\mathsf i}$.}
			\label{fig:IHC1}
		\end{figure}

		Denoting the matrix representation of the $SO(1,{ D})_\text{ext}$ part of a local representative \eqref{eq:localrepresentative} by $\mathbf g(x)$, it moves the point \eqref{eq:referencepoint} and the frames \eqref{eq:referenceframe} along the embedded $S^{{ D-2}}$:
		\begin{align}\label{eq:cosetmaps}
			X(x)={ \mathbf g}(x)\mathcal{X}_o\; ,\qquad
E_{\mathsf i}(x)={  \mathbf g(x)\mathbf S_{\mathsf i}\mathcal X_o}\; ,\qquad
N_{\mathsf a}(x)={  \mathbf g(x)\mathbf Z_{\mathsf a}\mathcal X_o}\;.
		\end{align}
		Since ${ \mathbf g}(x)\in SO(1,{ D})_\text{ext}$ preserves the ambient metric \eqref{eq:dS4embeddingspace}, the moving frames remain orthonormal:
		\begin{align}\label{eq:generalortho}
			{ E_{{\mathsf i}}}(x) \cdot { E_{{\mathsf j}}}(x) =\delta_{\mathsf i \mathsf j} \;, \qquad { E_{{\mathsf i}}}(x) \cdot { N_{{\mathsf a}}}(x) =0 \;, \qquad { N_{{\mathsf a}}}(x) \cdot { N_{{\mathsf b}}}(x)=\eta_{\mathsf a\mathsf b} \;.
		\end{align}
		However, at a general point, ${ E_{{\mathsf i}}}(x)$ (${ N_{{\mathsf a}}}(x)$) are not necessarily tangents (normals) at $X(x)$. In fact, taking the differential and using \eqref{eq:MCexpand}, we obtain
		\begin{align}\label{eq:dS4main-dX}
			dX={ \mathbf g}\left( { \mathbf g}^{-1}d{ \mathbf g}\right) \mathcal{X}_o  =
			{  \Theta^{\mathsf i}E_{\mathsf i}+\omega_Z^{\mathsf a}N_{\mathsf a}}\;,\qquad
			{ \Theta^{{\mathsf i}}}\coloneqq {  \omega_T^{\mathsf i}+\Delta\omega_S^{\mathsf i}}\; .
		\end{align}
		Thus, ${ \omega_Z^{\mathsf a}}$ are the physical normal components of the ambient displacement $dX$
		in the moving frame. They measure the failure of the chosen frame to be adapted to the embedded surface. The geometric meaning of the first IHC \eqref{eq:IHC1} is now clear: it requires the moving frame to be adapted to the surface so that $dX$ is purely tangent. See {Figure} \ref{fig:IHC1}.

		After imposing the first IHC \eqref{eq:IHC1}, we have 
		\begin{align}\label{eq:postIHCdiff}
			dX={ \mathbf g}\left( { \mathbf g}^{-1}d{ \mathbf g}\right) \mathcal{X}_o  =
			{ \Theta^{{\mathsf i}} \, E_{{\mathsf i}}} \; ,
		\end{align}
		so ${ \Theta^{{\mathsf i}}}$ is the induced vielbein on the embedded surface. From the definitions of \eqref{eq:omegaSexpand} and \eqref{eq:dS4main-dX}, this is related to the intrinsic vielbein \eqref{eq:veilbeinconnection} through
		\begin{align}
			{ \Theta^{{\mathsf i}}} = M^{\mathsf i}{}_{\mathsf j} { \bar\Theta^{{\mathsf j}}(x)} \;. 
		\end{align}
		We therefore have
		\begin{align}\label{eq:inducedintrinsicmetrics}
			M^T M = { \gamma^{-1}h} \;, \qquad h = { \delta_{{\mathsf i}{\mathsf j}}\Theta^{{\mathsf i}} \Theta^{{\mathsf j}}} \;\; \qquad \gamma = \delta_{\mathsf i \mathsf j}{ \bar\Theta^{{\mathsf i}} \bar\Theta^{{\mathsf j}}} \;,
		\end{align}
		where $\gamma$ is the intrinsic round $S^{{ D-2}}$ metric, while $h$ is the induced metric as embedded in $dS_{{ D}}$. 
		
		Note that any left multiplication $M\mapsto RM$ by a local ${ SO(D-2)}$ rotation would preserve the combination $\gamma^{-1}h$. The second IHC \eqref{eq:IHC2} eliminates this redundancy by demanding that $M$ is symmetric, 
		\begin{align}\label{eq:Msym}
			M^T = M \;.
		\end{align}
		With this condition, we can directly express $M$ in terms of the induced metric $h$ and the reference round metric $\gamma$. On the branch connected to the identity, {we have}
		\begin{align}\label{eq:MpostIHC}
			M= \sqrt{\gamma^{-1}h} \;.
		\end{align}

		\subsection{Geometric building blocks}
		
		As already mentioned, the MC coefficients \eqref{eq:veilbeinconnection} can be understood in terms of {the} intrinsic geometry of $S^{{ D-2}}_\text{int}$. Upon imposing the IHCs \eqref{eq:IHC1} and \eqref{eq:IHC2}, the MC components \eqref{eq:MCexpand} for the broken generators can similarly be identified as geometric objects.

		From \eqref{eq:veilbeinconnection}, \eqref{eq:dS4main-dX} and \eqref{eq:postIHCdiff}, we immediately see that upon the first IHC \eqref{eq:IHC1} $\Delta\omega_S^{\mathsf i}$ becomes the relative coframe comparing the induced and intrinsic coframes. With \eqref{eq:MpostIHC}, we can write
		\begin{align}
			{  \Delta\omega_S^{\mathsf i}} = { \mathsf S^{\mathsf i}{}_{\mathsf j}\bar\Theta^{{\mathsf j}}(x)}\;, \qquad { \mathsf S^{\mathsf i}{}_{\mathsf j}} \coloneqq { \left(\sqrt{\gamma^{-1}h}\right)^{{\mathsf i}}{}_{{\mathsf j}}}- \delta^{\mathsf i}{}_{\mathsf j} \;.
		\end{align}

		Similarly, ${\Delta\omega_J^{\mathsf i\mathsf j}}$ becomes the relative tangent-frame ${ SO(D-2)}$ connection upon imposing the first IHC \eqref{eq:IHC1}.
		The second IHC fixes the remaining local ${ SO(D-2)}$ freedom by choosing the polar, or symmetric-vielbein, representative \eqref{eq:MpostIHC}, so ${\Delta\omega_J^{\mathsf i\mathsf j}|_\text{IHC}}$ is the spin-connection difference in that chosen frame:
\begin{align}\label{eq:inducedspinconnection}
(\omega_h^\text{pol})^{\mathsf i\mathsf j}
=\omega^{\mathsf i\mathsf j}(x)+\Delta\omega_J^{\mathsf i\mathsf j}\big|_\text{IHC}\;,
\end{align}
		where ${(\omega_h^\text{pol})^{\mathsf i\mathsf j}}$ is the spin connection of the induced vielbein ${ \Theta^{{\mathsf i}}}$ in the symmetric-vielbein frame.

		Before imposing the first IHC \eqref{eq:IHC1}, differentiating the vectors ${ N_{{\mathsf a}}}(x)={  \mathbf g(x)\mathbf Z_{\mathsf a}\mathcal X_o}$ gives
		\begin{align}\label{eq:normalvariations}
			d{ N_{{\mathsf a}}}(x) & = { \mathbf g\left(\mathbf g^{-1}d\mathbf g\right)\mathbf Z_{\mathsf a}\mathcal X_o}
			={ -\frac1{\ell^2}\eta_{\mathsf a\mathsf b}\,\omega_Z^{\mathsf b}X}{ -}\eta_{\mathsf a\mathsf b}\,\omega^{\mathsf b{\mathsf i}}{ E_{{\mathsf i}}}
			{ +}\epsilon^{(\perp)}_{\mathsf a}{}^{\mathsf b}\,\omega_H { N_{{\mathsf b}}} \;.
		\end{align}
		After the first IHC \eqref{eq:IHC1}, the first term vanishes, while ${ E_{{\mathsf i}}}(x)$ and ${ N_{{\mathsf a}}}(x)$ become actual tangents and normals to the embedded surface. \eqref{eq:normalvariations} can then be understood in terms of standard submanifold geometry  ({see, e.g.,} \cite{Capovilla:1994deform}). First, the coefficient of the tangent vectors $E_{\mathsf i}$ encodes the second fundamental form, {i.e.,} extrinsic curvature, of the embedded surface, measured in the moving frame:
		\begin{align}
			\omega^{\mathsf a{\mathsf i}}\big|_\text{IHC}
			=-\eta^{\mathsf a\mathsf b}{ E_{{\mathsf i}}} \cdot d{ N_{{\mathsf b}}} \mathrel{\eqqcolon} { K^{{\mathsf a}{\mathsf i}}{}_{{\mathsf j}} \Theta^{{\mathsf j}}}
			={ K^{{\mathsf a}{\mathsf i}}{}_{{\mathsf j}}M^{{\mathsf j}}{}_{{\mathsf k}}\bar\Theta^{{\mathsf k}}(x)} \;,
			\qquad K^{\mathsf a}{}_{ij}=K^{\mathsf a}{}_{ji}\;.
		\end{align}
		Second, the coefficient of the normal vectors $N_{\mathsf a}$, or explicitly
			\begin{align}
				\omega_H\big|_\text{IHC}
				=\frac12 \epsilon^{(\perp)}{}^{{ \mathsf a \mathsf b}} { N_{{\mathsf a}}}\cdot d{ N_{{\mathsf b}}}
				\mathrel{\eqqcolon}{ {\mathcal A_{{\mathsf i}}}\Theta^{{\mathsf i}}}
				={ {\mathcal A_{{\mathsf i}}}M^{{\mathsf i}}{}_{{\mathsf j}}\bar\Theta^{{\mathsf j}}(x)} \;,
			\end{align}
			is the $SO(1,1)$ connection on the rank-two Lorentzian normal bundle expanded in the moving frame, which records the infinitesimal \(SO(1,1)\) boost of the chosen normal frame as one moves along the surface.

		To summarize, the building blocks for {the EFTs} associated with the symmetry breaking pattern \eqref{eq:edgesymbreak} include 
\begin{align}\label{eq:cosetGeometricBlocks}
\bar\Theta^{\mathsf i}\;,\qquad\omega^{\mathsf i\mathsf j}\;,\qquad
\mathsf S^{\mathsf i}{}_{\mathsf j}\bar\Theta^{\mathsf j}\;, \qquad \Delta\omega_J^{\mathsf i\mathsf j}\big|_\text{IHC} \;, \qquad 
K^{\mathsf a\mathsf i}{}_{\mathsf j}M^{\mathsf j}{}_{\mathsf k}\bar\Theta^{\mathsf k}\;,\qquad
{\mathcal A_{\mathsf i}}M^{\mathsf i}{}_{\mathsf j}\bar\Theta^{\mathsf j}\;.
\end{align}
	These building blocks are equivalent to those in \eqref{eq:generalEFT}, expressed in an orthonormal frame. They transform according to \eqref{eq:MCcomponentTransformations} and \eqref{eq:MCcomponentTransformationsunbroken}, as expected for these geometric quantities. As in the direct geometric construction in {Section} \ref{sec:EFT}, {we restrict to strictly invariant local densities constructed from} \eqref{eq:cosetGeometricBlocks} and their covariant derivatives, and leave possible Wess-Zumino terms for future investigation.

		\bibliographystyle{utphys}
		\bibliography{ref}

@article{Agarwal:2016cir,
    author = {Agarwal, Abhishek and Karabali, Dimitra and Nair, V. P.},
    title = {{Gauge-invariant Variables and Entanglement Entropy}},
    journal = {Phys. Rev. D},
    volume = {96},
    number = {12},
    pages = {125008},
    year = {2017},
    doi = {10.1103/PhysRevD.96.125008},
    eprint = {1701.00014},
    archivePrefix = {arXiv},
    primaryClass = {hep-th}
}

@article{Aguilar-Gutierrez:2026svf,
    author = {Aguilar-Gutierrez, Sergio E. and Ferrero, Renata and Hoehn, Philipp A. and Marchetti, Luca},
    title = {{Relational path integral, effective actions and quantum frame covariance in gravity}},
    year = {2026},
    month = {7},
    eprint = {2607.21463},
    archivePrefix = {arXiv},
    primaryClass = {hep-th}
}

@article{Anninos:2011af,
    author = {Anninos, Dionysios and Hartnoll, Sean A. and Hofman, Diego M.},
    title = {{Static Patch Solipsism: Conformal Symmetry of the de Sitter Worldline}},
    journal = {Class. Quant. Grav.},
    volume = {29},
    pages = {075002},
    year = {2012},
    doi = {10.1088/0264-9381/29/7/075002},
    eprint = {1109.4942},
    archivePrefix = {arXiv},
    primaryClass = {hep-th}
}

@article{Anninos:2011zn,
    author = {Anninos, Dionysios and Anous, Tarek and Bredberg, Irene and Ng, Gim Seng},
    title = {{Incompressible Fluids of the de Sitter Horizon and Beyond}},
    journal = {JHEP},
    volume = {05},
    pages = {107},
    year = {2012},
    doi = {10.1007/JHEP05(2012)107},
    eprint = {1110.3792},
    archivePrefix = {arXiv},
    primaryClass = {hep-th}
}

@article{Anninos:2020hfj,
    author = {Anninos, Dionysios and Denef, Frederik and Law, Y. T. Albert and Sun, Zimo},
    title = {{Quantum de Sitter horizon entropy from quasicanonical bulk, edge, sphere and topological string partition functions}},
    journal = {JHEP},
    volume = {01},
    pages = {088},
    year = {2022},
    doi = {10.1007/JHEP01(2022)088},
    eprint = {2009.12464},
    archivePrefix = {arXiv},
    primaryClass = {hep-th}
}

@article{Anninos:2021ihe,
    author = {Anninos, Dionysios and Harris, Eleanor},
    title = {{Three-dimensional de Sitter horizon thermodynamics}},
    journal = {JHEP},
    volume = {10},
    pages = {091},
    year = {2021},
    doi = {10.1007/JHEP10(2021)091},
    eprint = {2106.13832},
    archivePrefix = {arXiv},
    primaryClass = {hep-th}
}

@article{Anninos:2024wpy,
    author = {Anninos, Dionysios and Galante, Dami\'an A. and Maneerat, Chawakorn},
    title = {{Cosmological observatories}},
    journal = {Class. Quant. Grav.},
    volume = {41},
    number = {16},
    pages = {165009},
    year = {2024},
    doi = {10.1088/1361-6382/ad5824},
    eprint = {2402.04305},
    archivePrefix = {arXiv},
    primaryClass = {hep-th}
}

@article{Anninos:2025zgr,
    author = {Anninos, Dionysios and Galante, Dami{\'a}n A. and Georgescu, Silvia and Maneerat, Chawakorn and Svesko, Andrew},
    title = {{The stretched horizon limit}},
    journal = {JHEP},
    volume = {05},
    pages = {234},
    year = {2026},
    doi = {10.1007/JHEP05(2026)234},
    eprint = {2512.16738},
    archivePrefix = {arXiv},
    primaryClass = {hep-th}
}

@article{Araujo-Regado:2024dpr,
    author = "Araujo-Regado, Goncalo and Hoehn, Philipp A. and Sartini, Francesco and Tomova, Bilyana",
    title = "{Soft edges: the many links between soft and edge modes}",
    eprint = "2412.14548",
    archivePrefix = "arXiv",
    primaryClass = "hep-th",
    doi = "10.1007/JHEP07(2025)180",
    journal = "JHEP",
    volume = "07",
    pages = "180",
    year = "2025"
}

@article{Balachandran:1994EdgeStates,
    author = {A. P. Balachandran and L. Chandar and Arshad Momen},
    title = {{Edge states in Gravity and Black Hole Physics}},
    journal = {Nucl. Phys. B},
    volume = {461},
    pages = {581--596},
    year = {1996},
    doi = {10.1016/0550-3213(95)00622-2},
    eprint = {gr-qc/9412019},
    archivePrefix = {arXiv}
}

@article{Balasubramanian:2023dpj,
    author = "Balasubramanian, Vijay and Cummings, Charlie",
    title = "{The entropy of finite gravitating regions}",
    eprint = "2312.08434",
    archivePrefix = "arXiv",
    primaryClass = "hep-th",
    month = "12",
    year = "2023"
}

@article{Ball:2024gti,
    author = "Ball, Adam and Ciambelli, Luca",
    title = "{Dynamical edge modes in Yang-Mills theory}",
    eprint = "2412.06672",
    archivePrefix = "arXiv",
    primaryClass = "hep-th",
    doi = "10.21468/SciPostPhys.20.1.013",
    journal = "SciPost Phys.",
    volume = "20",
    number = "1",
    pages = "013",
    year = "2026"
}

@article{Ball:2024hqe,
    author = "Ball, Adam and Law, Y. T. Albert and Wong, Gabriel",
    title = "{Dynamical edge modes and entanglement in Maxwell theory}",
    eprint = "2403.14542",
    archivePrefix = "arXiv",
    primaryClass = "hep-th",
    doi = "10.1007/JHEP09(2024)032",
    journal = "JHEP",
    volume = "09",
    pages = "032",
    year = "2024"
}

@article{Ball:2024xhf,
    author = "Ball, Adam and Law, Y. T. Albert",
    title = "{Dynamical edge modes in p-form gauge theories}",
    eprint = "2411.02555",
    archivePrefix = "arXiv",
    primaryClass = "hep-th",
    doi = "10.1007/JHEP02(2025)182",
    journal = "JHEP",
    volume = "02",
    pages = "182",
    year = "2025"
}

@inproceedings{Banks:2003cg,
    author = {Banks, T.},
    title = {{Some thoughts on the quantum theory of de sitter space}},
    booktitle = {{The Davis Meeting on Cosmic Inflation}},
    year = {2003},
    month = {5},
    eprint = {astro-ph/0305037},
    archivePrefix = {arXiv},
    reportnumber = {DAVISINFLATION-2003-BANKS}
}

@article{Banks:2006rx,
    author = {Banks, Tom and Fiol, Bartomeu and Morisse, Alexander},
    title = {{Towards a quantum theory of de Sitter space}},
    journal = {JHEP},
    volume = {12},
    pages = {004},
    year = {2006},
    doi = {10.1088/1126-6708/2006/12/004},
    eprint = {hep-th/0609062},
    archivePrefix = {arXiv},
    reportnumber = {RUNHETC-06-21, SCIPP-06-11, ITFA-2006-31}
}

@article{Banks:2013qpa,
    author = {Banks, Tom},
    title = {{Lectures on Holographic Space Time}},
    year = {2013},
    month = {11},
    eprint = {1311.0755},
    archivePrefix = {arXiv},
    primaryClass = {hep-th},
    reportnumber = {RUNHETC-2013-1-SCIPP-13-10}
}

@article{Barnich:2018zdg,
    author = {Barnich, Glenn},
    title = {{Black hole entropy from nonproper gauge degrees of freedom: The charged vacuum capacitor}},
    journal = {Phys. Rev. D},
    volume = {99},
    number = {2},
    pages = {026007},
    year = {2019},
    doi = {10.1103/PhysRevD.99.026007},
    eprint = {1806.00549},
    archivePrefix = {arXiv},
    primaryClass = {hep-th}
}

@article{Benedetti:2019uej,
    author = {Benedetti, Valentin and Casini, Horacio},
    title = {{Entanglement entropy of linearized gravitons in a sphere}},
    journal = {Phys. Rev. D},
    volume = {101},
    number = {4},
    pages = {045004},
    year = {2020},
    doi = {10.1103/PhysRevD.101.045004},
    eprint = {1908.01800},
    archivePrefix = {arXiv},
    primaryClass = {hep-th}
}

@article{Bhattacharyya:2007vjd,
    author = {Bhattacharyya, Sayantani and Hubeny, Veronika E and Minwalla, Shiraz and Rangamani, Mukund},
    title = {{Nonlinear Fluid Dynamics from Gravity}},
    journal = {JHEP},
    volume = {02},
    pages = {045},
    year = {2008},
    doi = {10.1088/1126-6708/2008/02/045},
    eprint = {0712.2456},
    archivePrefix = {arXiv},
    primaryClass = {hep-th},
    reportnumber = {TIFR-TH-07-44, DCPT-07-73, NI07097}
}

@article{blommaert2025gravitons,
    author = "Blommaert, Andreas and Colin-Ellerin, Sean",
    title = "{Gravitons on the edge}",
    eprint = "2405.12276",
    archivePrefix = "arXiv",
    primaryClass = "hep-th",
    doi = "10.1007/JHEP03(2025)116",
    journal = "JHEP",
    volume = "03",
    pages = "116",
    year = "2025"
}

@article{Blommaert:2018oue,
    author = {Blommaert, Andreas and Mertens, Thomas G. and Verschelde, Henri},
    title = {{Edge dynamics from the path integral \textemdash{} Maxwell and Yang-Mills}},
    journal = {JHEP},
    volume = {11},
    pages = {080},
    year = {2018},
    doi = {10.1007/JHEP11(2018)080},
    eprint = {1804.07585},
    archivePrefix = {arXiv},
    primaryClass = {hep-th}
}

@article{Blommaert:2018rsf,
    author = {Blommaert, Andreas and Mertens, Thomas G. and Verschelde, Henri and Zakharov, Valentin I.},
    title = {{Edge State Quantization: Vector Fields in Rindler}},
    journal = {JHEP},
    volume = {08},
    pages = {196},
    year = {2018},
    doi = {10.1007/JHEP08(2018)196},
    eprint = {1801.09910},
    archivePrefix = {arXiv},
    primaryClass = {hep-th}
}

@article{Bonifacio:2018zex,
    author = {Bonifacio, James and Hinterbichler, Kurt and Joyce, Austin and Rosen, Rachel A.},
    title = {{Shift Symmetries in (Anti) de Sitter Space}},
    journal = {JHEP},
    volume = {02},
    pages = {178},
    year = {2019},
    doi = {10.1007/JHEP02(2019)178},
    eprint = {1812.08167},
    archivePrefix = {arXiv},
    primaryClass = {hep-th}
}

@article{Bonifacio:2019hrj,
    author = {Bonifacio, James and Hinterbichler, Kurt and Johnson, Laura A. and Joyce, Austin},
    title = {{Shift-Symmetric Spin-1 Theories}},
    journal = {JHEP},
    volume = {09},
    pages = {029},
    year = {2019},
    doi = {10.1007/JHEP09(2019)029},
    eprint = {1906.10692},
    archivePrefix = {arXiv},
    primaryClass = {hep-th}
}

@article{Bonifacio:2021mrf,
    author = {Bonifacio, James and Hinterbichler, Kurt and Joyce, Austin and Roest, Diederik},
    title = {{Exceptional scalar theories in de Sitter space}},
    journal = {JHEP},
    volume = {04},
    pages = {128},
    year = {2022},
    doi = {10.1007/JHEP04(2022)128},
    eprint = {2112.12151},
    archivePrefix = {arXiv},
    primaryClass = {hep-th}
}

@article{Boyda:2002ba,
    author = {Boyda, Edward K. and Ganguli, Surya and Horava, Petr and Varadarajan, Uday},
    title = {{Holographic protection of chronology in universes of the Godel type}},
    eprint = {hep-th/0212087},
    archivePrefix = {arXiv},
    reportnumber = {LBNL-51811, UCB-PTH-02-57},
    doi = {10.1103/PhysRevD.67.106003},
    journal = {Phys. Rev. D},
    volume = {67},
    pages = {106003},
    year = {2003}
}

@article{Brauner:2014aha,
    author = {Brauner, Tom{\'a}{\v{s}} and Watanabe, Haruki},
    title = {{Spontaneous breaking of spacetime symmetries and the inverse Higgs effect}},
    journal = {Phys. Rev. D},
    volume = {89},
    number = {8},
    pages = {085004},
    year = {2014},
    doi = {10.1103/PhysRevD.89.085004},
    eprint = {1401.5596},
    archivePrefix = {arXiv},
    primaryClass = {hep-ph}
}

@article{Bredberg:2011jq,
    author = {Bredberg, Irene and Keeler, Cynthia and Lysov, Vyacheslav and Strominger, Andrew},
    title = {{From Navier-Stokes To Einstein}},
    journal = {JHEP},
    volume = {07},
    pages = {146},
    year = {2012},
    doi = {10.1007/JHEP07(2012)146},
    eprint = {1101.2451},
    archivePrefix = {arXiv},
    primaryClass = {hep-th}
}

@article{Callan:1969sn,
    author = {Callan, Jr., Curtis G. and Coleman, Sidney R. and Wess, J. and Zumino, Bruno},
    title = {{Structure of phenomenological Lagrangians. 2.}},
    journal = {Phys. Rev.},
    volume = {177},
    pages = {2247--2250},
    year = {1969},
    doi = {10.1103/PhysRev.177.2247}
}

@article{Capovilla:1994deform,
    author = {Capovilla, Riccardo and Guven, Jemal},
    title = {{Geometry of deformations of relativistic membranes}},
    journal = {Phys. Rev. D},
    volume = {51},
    pages = {6736--6743},
    year = {1995},
    doi = {10.1103/PhysRevD.51.6736},
    eprint = {gr-qc/9411060},
    archivePrefix = {arXiv},
    reportnumber = {CIEA-GR-9407, ICN-UNAM-9405}
}

@article{Carrozza:2021gju,
    author = {Carrozza, Sylvain and Hoehn, Philipp A.},
    title = {{Edge modes as reference frames and boundary actions from post-selection}},
    journal = {JHEP},
    volume = {02},
    pages = {172},
    year = {2022},
    doi = {10.1007/JHEP02(2022)172},
    eprint = {2109.06184},
    archivePrefix = {arXiv},
    primaryClass = {hep-th}
}

@article{Carrozza:2022xut,
    author = {Carrozza, Sylvain and Eccles, Stefan and Hoehn, Philipp A.},
    title = {{Edge modes as dynamical frames: charges from post-selection in generally covariant theories}},
    journal = {SciPost Phys.},
    volume = {17},
    number = {2},
    pages = {048},
    year = {2024},
    doi = {10.21468/SciPostPhys.17.2.048},
    eprint = {2205.00913},
    archivePrefix = {arXiv},
    primaryClass = {hep-th}
}

@article{Casini:2013rba,
    author = {Casini, Horacio and Huerta, Marina and Rosabal, Jose Alejandro},
    title = {{Remarks on entanglement entropy for gauge fields}},
    journal = {Phys. Rev. D},
    volume = {89},
    number = {8},
    pages = {085012},
    year = {2014},
    doi = {10.1103/PhysRevD.89.085012},
    eprint = {1312.1183},
    archivePrefix = {arXiv},
    primaryClass = {hep-th}
}

@article{Chandrasekaran:2022cip,
    author = "Chandrasekaran, Venkatesa and Longo, Roberto and Penington, Geoff and Witten, Edward",
    title = "{An algebra of observables for de Sitter space}",
    eprint = "2206.10780",
    archivePrefix = "arXiv",
    primaryClass = "hep-th",
    doi = "10.1007/JHEP02(2023)082",
    journal = "JHEP",
    volume = "02",
    pages = "082",
    year = "2023"
}

@article{Chandrasekaran:2026pnc,
    author = "Chandrasekaran, Venkatesa and Flanagan, {\'E}anna {\'E}.",
    title = "{Subregion algebras in classical and quantum gravity}",
    eprint = "2601.07915",
    archivePrefix = "arXiv",
    primaryClass = "hep-th",
    month = "1",
    year = "2026"
}

@article{Chen:2024rpx,
    author = "Chen, Chang-Han and Penington, Geoff",
    title = "{A clock is just a way to tell the time: gravitational algebras in cosmological spacetimes}",
    eprint = "2406.02116",
    archivePrefix = "arXiv",
    primaryClass = "hep-th",
    month = "6",
    year = "2024"
}

@article{Chen:2025jqm,
    author = {Chen, Yiming and Stanford, Douglas and Tang, Haifeng and Yang, Zhenbin},
    title = {{On the phase of the de Sitter density of states}},
    eprint = {2511.01400},
    archivePrefix = {arXiv},
    primaryClass = {hep-th},
    doi = {10.1007/JHEP05(2026)068},
    journal = {JHEP},
    volume = {05},
    pages = {068},
    year = {2026}
}

@article{Ciambelli:2021nmv,
    author = "Ciambelli, Luca and Leigh, Robert G. and Pai, Pin-Chun",
    title = "{Embeddings and Integrable Charges for Extended Corner Symmetry}",
    eprint = "2111.13181",
    archivePrefix = "arXiv",
    primaryClass = "hep-th",
    doi = "10.1103/PhysRevLett.128.171302",
    journal = "Phys. Rev. Lett.",
    volume = "128",
    year = "2022",
    pages = "171302"
}

@article{Ciambelli:2021vnn,
    author = {Ciambelli, Luca and Leigh, Robert G.},
    title = {{Isolated surfaces and symmetries of gravity}},
    journal = {Phys. Rev. D},
    volume = {104},
    number = {4},
    pages = {046005},
    year = {2021},
    doi = {10.1103/PhysRevD.104.046005},
    eprint = {2104.07643},
    archivePrefix = {arXiv},
    primaryClass = {hep-th}
}

@article{Ciambelli:2022cfr,
    author = {Ciambelli, Luca and Leigh, Robert G.},
    title = {{Universal corner symmetry and the orbit method for gravity}},
    journal = {Nucl. Phys. B},
    volume = {986},
    pages = {116053},
    year = {2023},
    doi = {10.1016/j.nuclphysb.2022.116053},
    eprint = {2207.06441},
    archivePrefix = {arXiv},
    primaryClass = {hep-th}
}

@article{Ciambelli:2022vot,
    author = {Ciambelli, Luca},
    title = {{From Asymptotic Symmetries to the Corner Proposal}},
    journal = {PoS},
    volume = {Modave2022},
    pages = {002},
    year = {2023},
    doi = {10.22323/1.435.0002},
    eprint = {2212.13644},
    archivePrefix = {arXiv},
    primaryClass = {hep-th}
}

@article{Clark:2005ht,
    author = {Clark, T. E. and Love, S. T. and Nitta, Muneto and ter Veldhuis, T.},
    title = {{AdS(d+1) ---\ensuremath{>} AdS(d)}},
    journal = {J. Math. Phys.},
    volume = {46},
    pages = {102304},
    year = {2005},
    doi = {10.1063/1.2048307},
    eprint = {hep-th/0501241},
    archivePrefix = {arXiv},
    reportnumber = {FERMILAB-PUB-05-073-T}
}

@article{Clark:2006pd,
    author = {Clark, T. E. and Love, S. T. and Nitta, Muneto and ter Veldhuis, T. and Xiong, Chi},
    title = {{Gravitating p-Branes}},
    journal = {Phys. Rev. D},
    volume = {75},
    pages = {065028},
    year = {2007},
    doi = {10.1103/PhysRevD.75.065028},
    eprint = {hep-th/0612147},
    archivePrefix = {arXiv}
}

@article{Clark:2007rn,
    author = {Clark, T. E. and Love, S. T. and Nitta, Muneto and ter Veldhuis, T. and Xiong, Chi},
    title = {{Oscillating p-branes}},
    journal = {Phys. Rev. D},
    volume = {76},
    pages = {105014},
    year = {2007},
    doi = {10.1103/PhysRevD.76.105014},
    eprint = {hep-th/0703179},
    archivePrefix = {arXiv}
}

@article{Coleman:1969sm,
    author = {Coleman, Sidney R. and Wess, J. and Zumino, Bruno},
    title = {{Structure of phenomenological Lagrangians. 1.}},
    journal = {Phys. Rev.},
    volume = {177},
    pages = {2239--2247},
    year = {1969},
    doi = {10.1103/PhysRev.177.2239}
}

@article{Dabholkar:1994ai,
    author = {Dabholkar, Atish},
    title = {{Strings on a cone and black hole entropy}},
    editor = {Das, Sumit R. and Mandal, Gautam and Mukhi, S. and Wadia, S. R.},
    journal = {Nucl. Phys. B},
    volume = {439},
    pages = {650--664},
    year = {1995},
    doi = {10.1016/0550-3213(95)00050-3},
    eprint = {hep-th/9408098},
    archivePrefix = {arXiv},
    reportnumber = {HUTP-94-A019}
}

@article{Dabholkar:2023ows,
    author = {Dabholkar, Atish and Moitra, Upamanyu},
    title = {{Finite entanglement entropy in string theory}},
    eprint = {2306.00990},
    archivePrefix = {arXiv},
    primaryClass = {hep-th},
    doi = {10.1103/PhysRevD.109.L121901},
    journal = {Phys. Rev. D},
    volume = {109},
    number = {12},
    pages = {L121901},
    year = {2024}
}

@article{Dabholkar:2023tzd,
    author = {Dabholkar, Atish and Moitra, Upamanyu},
    title = {{Quantum entanglement on black hole horizons in string theory and holography}},
    journal = {JHEP},
    volume = {06},
    pages = {053},
    year = {2024},
    doi = {10.1007/JHEP06(2024)053},
    eprint = {2312.14253},
    archivePrefix = {arXiv},
    primaryClass = {hep-th}
}

@article{Dabholkar:2024neq,
    author = "Dabholkar, Atish and Moitra, Upamanyu",
    title = "{Heterotic strings and quantum entanglement}",
    eprint = "2407.17553",
    archivePrefix = "arXiv",
    primaryClass = "hep-th",
    doi = "10.1007/JHEP12(2024)012",
    journal = "JHEP",
    volume = "12",
    pages = "012",
    year = "2024"
}

@article{David:2022jfd,
    author = {David, Justin R. and Mukherjee, Jyotirmoy},
    title = {{Entanglement entropy of gravitational edge modes}},
    journal = {JHEP},
    volume = {08},
    pages = {065},
    year = {2022},
    doi = {10.1007/JHEP08(2022)065},
    eprint = {2201.06043},
    archivePrefix = {arXiv},
    primaryClass = {hep-th}
}

@article{Delcamp:2016eya,
    author = {Delcamp, Clement and Dittrich, Bianca and Riello, Aldo},
    title = {{On entanglement entropy in non-Abelian lattice gauge theory and 3D quantum gravity}},
    journal = {JHEP},
    volume = {11},
    pages = {102},
    year = {2016},
    doi = {10.1007/JHEP11(2016)102},
    eprint = {1609.04806},
    archivePrefix = {arXiv},
    primaryClass = {hep-th}
}

@article{DeVuyst:2024fxc,
    author = "De Vuyst, Julian and Eccles, Stefan and Hoehn, Philipp A. and Kirklin, Josh",
    title = "{Crossed products and quantum reference frames: on the observer-dependence of gravitational entropy}",
    eprint = "2412.15502",
    archivePrefix = "arXiv",
    primaryClass = "hep-th",
    doi = "10.1007/JHEP07(2025)063",
    journal = "JHEP",
    volume = "07",
    pages = "063",
    year = "2025",
    note = "[Erratum: JHEP 10, 234 (2025)]"
}

@article{DeVuyst:2024khu,
    author = "De Vuyst, Julian and Eccles, Stefan and Hoehn, Philipp A. and Kirklin, Josh",
    title = "{Gravitational entropy is observer-dependent}",
    eprint = "2405.00114",
    archivePrefix = "arXiv",
    primaryClass = "hep-th",
    doi = "10.1007/JHEP07(2025)146",
    journal = "JHEP",
    volume = "07",
    pages = "146",
    year = "2025"
}

@article{Donnelly:2011hn,
    author = {Donnelly, William},
    title = {{Decomposition of entanglement entropy in lattice gauge theory}},
    journal = {Phys. Rev. D},
    volume = {85},
    pages = {085004},
    year = {2012},
    doi = {10.1103/PhysRevD.85.085004},
    eprint = {1109.0036},
    archivePrefix = {arXiv},
    primaryClass = {hep-th}
}

@article{Donnelly:2012st,
    author = {Donnelly, William and Wall, Aron C.},
    title = {{Do gauge fields really contribute negatively to black hole entropy?}},
    journal = {Phys. Rev. D},
    volume = {86},
    pages = {064042},
    year = {2012},
    doi = {10.1103/PhysRevD.86.064042},
    eprint = {1206.5831},
    archivePrefix = {arXiv},
    primaryClass = {hep-th}
}

@article{Donnelly:2014fua,
    author = {Donnelly, William and Wall, Aron C.},
    title = {{Entanglement entropy of electromagnetic edge modes}},
    journal = {Phys. Rev. Lett.},
    volume = {114},
    number = {11},
    pages = {111603},
    year = {2015},
    doi = {10.1103/PhysRevLett.114.111603},
    eprint = {1412.1895},
    archivePrefix = {arXiv},
    primaryClass = {hep-th}
}

@article{Donnelly:2014gva,
    author = {Donnelly, William},
    title = {{Entanglement entropy and nonabelian gauge symmetry}},
    journal = {Class. Quant. Grav.},
    volume = {31},
    number = {21},
    pages = {214003},
    year = {2014},
    doi = {10.1088/0264-9381/31/21/214003},
    eprint = {1406.7304},
    archivePrefix = {arXiv},
    primaryClass = {hep-th}
}

@article{Donnelly:2015hxa,
    author = {Donnelly, William and Wall, Aron C.},
    title = {{Geometric entropy and edge modes of the electromagnetic field}},
    journal = {Phys. Rev. D},
    volume = {94},
    number = {10},
    pages = {104053},
    year = {2016},
    doi = {10.1103/PhysRevD.94.104053},
    eprint = {1506.05792},
    archivePrefix = {arXiv},
    primaryClass = {hep-th}
}

@article{Donnelly:2016auv,
    author = {Donnelly, William and Freidel, Laurent},
    title = {{Local subsystems in gauge theory and gravity}},
    journal = {JHEP},
    volume = {09},
    pages = {102},
    year = {2016},
    doi = {10.1007/JHEP09(2016)102},
    eprint = {1601.04744},
    archivePrefix = {arXiv},
    primaryClass = {hep-th}
}

@article{Donnelly:2016jet,
    author = {Donnelly, William and Wong, Gabriel},
    title = {{Entanglement branes in a two-dimensional string theory}},
    journal = {JHEP},
    volume = {09},
    pages = {097},
    year = {2017},
    doi = {10.1007/JHEP09(2017)097},
    eprint = {1610.01719},
    archivePrefix = {arXiv},
    primaryClass = {hep-th}
}

@article{Donnelly:2018ppr,
    author = {Donnelly, William and Wong, Gabriel},
    title = {{Entanglement branes, modular flow, and extended topological quantum field theory}},
    journal = {JHEP},
    volume = {10},
    pages = {016},
    year = {2019},
    doi = {10.1007/JHEP10(2019)016},
    eprint = {1811.10785},
    archivePrefix = {arXiv},
    primaryClass = {hep-th}
}

@article{Donnelly:2020teo,
    author = {Donnelly, William and Jiang, Yikun and Kim, Manki and Wong, Gabriel},
    title = {{Entanglement entropy and edge modes in topological string theory. Part I. Generalized entropy for closed strings}},
    journal = {JHEP},
    volume = {10},
    pages = {201},
    year = {2021},
    doi = {10.1007/JHEP10(2021)201},
    eprint = {2010.15737},
    archivePrefix = {arXiv},
    primaryClass = {hep-th}
}

@article{Donnelly:2020xgu,
    author = {Donnelly, William and Freidel, Laurent and Moosavian, Seyed Faroogh and Speranza, Antony J.},
    title = {{Gravitational edge modes, coadjoint orbits, and hydrodynamics}},
    journal = {JHEP},
    volume = {09},
    pages = {008},
    year = {2021},
    doi = {10.1007/JHEP09(2021)008},
    eprint = {2012.10367},
    archivePrefix = {arXiv},
    primaryClass = {hep-th}
}

@article{Donnelly:2022kfs,
    author = {Donnelly, William and Freidel, Laurent and Moosavian, Seyed Faroogh and Speranza, Antony J.},
    title = {{Matrix Quantization of Gravitational Edge Modes}},
    journal = {JHEP},
    volume = {05},
    pages = {163},
    year = {2023},
    doi = {10.1007/JHEP05(2023)163},
    eprint = {2212.09120},
    archivePrefix = {arXiv},
    primaryClass = {hep-th}
}

@article{Eling:2016xlx,
    author = {Eling, Christopher and Oz, Yaron},
    title = {{On the Membrane Paradigm and Spontaneous Breaking of Horizon BMS Symmetries}},
    journal = {JHEP},
    volume = {07},
    pages = {065},
    year = {2016},
    doi = {10.1007/JHEP07(2016)065},
    eprint = {1605.00183},
    archivePrefix = {arXiv},
    primaryClass = {hep-th}
}

@article{Esposito:2026diu,
    author = {Esposito, Angelo and Nicolis, Alberto and Penco, Riccardo},
    title = {{Effective Field Theories for Material Media}},
    year = {2026},
    month = {7},
    eprint = {2607.06666},
    archivePrefix = {arXiv},
    primaryClass = {hep-th}
}

@article{fewster2025semi,
    author = {Fewster, Christopher J and Janssen, Daan W and Rejzner, Kasia},
    title = {Semi-local observables, edge modes and quantum reference frames in quantum electromagnetism: an algebraic approach},
    eprint = {2508.20939},
    archivePrefix = {arXiv},
    primaryClass = {math-ph},
    note = {Accepted for publication in Commun. Math. Phys.},
    year = {2025}
}

@article{Freidel:2018fsk,
    author = {Freidel, Laurent and Pranzetti, Daniele},
    title = {{Electromagnetic duality and central charge}},
    journal = {Phys. Rev. D},
    volume = {98},
    number = {11},
    pages = {116008},
    year = {2018},
    doi = {10.1103/PhysRevD.98.116008},
    eprint = {1806.03161},
    archivePrefix = {arXiv},
    primaryClass = {hep-th}
}

@article{Freidel:2019ees,
    author = {Freidel, Laurent and Livine, Etera R. and Pranzetti, Daniele},
    title = {{Gravitational edge modes: from Kac\textendash{}Moody charges to Poincar\'e networks}},
    journal = {Class. Quant. Grav.},
    volume = {36},
    number = {19},
    pages = {195014},
    year = {2019},
    doi = {10.1088/1361-6382/ab40fe},
    eprint = {1906.07876},
    archivePrefix = {arXiv},
    primaryClass = {hep-th}
}

@article{Freidel:2020ayo,
    author = {Freidel, Laurent and Geiller, Marc and Pranzetti, Daniele},
    title = {{Edge modes of gravity. Part III. Corner simplicity constraints}},
    journal = {JHEP},
    volume = {01},
    pages = {100},
    year = {2021},
    doi = {10.1007/JHEP01(2021)100},
    eprint = {2007.12635},
    archivePrefix = {arXiv},
    primaryClass = {hep-th}
}

@article{Freidel:2020svx,
    author = {Freidel, Laurent and Geiller, Marc and Pranzetti, Daniele},
    title = {{Edge modes of gravity. Part II. Corner metric and Lorentz charges}},
    journal = {JHEP},
    volume = {11},
    pages = {027},
    year = {2020},
    doi = {10.1007/JHEP11(2020)027},
    eprint = {2007.03563},
    archivePrefix = {arXiv},
    primaryClass = {hep-th}
}

@article{Freidel:2020xyx,
    author = {Freidel, Laurent and Geiller, Marc and Pranzetti, Daniele},
    title = {{Edge modes of gravity. Part I. Corner potentials and charges}},
    journal = {JHEP},
    volume = {11},
    pages = {026},
    year = {2020},
    doi = {10.1007/JHEP11(2020)026},
    eprint = {2006.12527},
    archivePrefix = {arXiv},
    primaryClass = {hep-th}
}

@article{Freidel:2021cjp,
    author = {Freidel, Laurent and Oliveri, Roberto and Pranzetti, Daniele and Speziale, Simone},
    title = {{Extended corner symmetry, charge bracket and Einstein\textquoteright{}s equations}},
    journal = {JHEP},
    volume = {09},
    pages = {083},
    year = {2021},
    doi = {10.1007/JHEP09(2021)083},
    eprint = {2104.12881},
    archivePrefix = {arXiv},
    primaryClass = {hep-th}
}

@article{Freidel:2021dxw,
    author = {Freidel, Laurent},
    title = {{A canonical bracket for open gravitational system}},
    year = {2021},
    month = {11},
    eprint = {2111.14747},
    archivePrefix = {arXiv},
    primaryClass = {hep-th}
}

@article{Freidel:2025AnomalousFrames,
    author = "Freidel, Laurent and Kirklin, Josh",
    title = "{Localization and anomalous reference frames in gravity}",
    eprint = "2510.26589",
    archivePrefix = "arXiv",
    primaryClass = "hep-th",
    doi = "10.1103/9wk3-jfzn",
    journal = "Phys. Rev. D",
    volume = "114",
    number = "4",
    pages = "046017",
    year = "2026"
}

@article{Gabadadze:2012tr,
    author = {Gabadadze, Gregory and Hinterbichler, Kurt and Khoury, Justin and Pirtskhalava, David and Trodden, Mark},
    title = {{A Covariant Master Theory for Novel Galilean Invariant Models and Massive Gravity}},
    journal = {Phys. Rev. D},
    volume = {86},
    pages = {124004},
    year = {2012},
    doi = {10.1103/PhysRevD.86.124004},
    eprint = {1208.5773},
    archivePrefix = {arXiv},
    primaryClass = {hep-th},
    reportnumber = {UCSD-PTH-12-11}
}

@article{Geiller:2017whh,
    author = {Geiller, Marc},
    title = {{Lorentz-diffeomorphism edge modes in 3d gravity}},
    journal = {JHEP},
    volume = {02},
    pages = {029},
    year = {2018},
    doi = {10.1007/JHEP02(2018)029},
    eprint = {1712.05269},
    archivePrefix = {arXiv},
    primaryClass = {gr-qc}
}

@article{Geiller:2017xad,
    author = {Geiller, Marc},
    title = {{Edge modes and corner ambiguities in 3d Chern\textendash{}Simons theory and gravity}},
    journal = {Nucl. Phys. B},
    volume = {924},
    pages = {312--365},
    year = {2017},
    doi = {10.1016/j.nuclphysb.2017.09.010},
    eprint = {1703.04748},
    archivePrefix = {arXiv},
    primaryClass = {gr-qc}
}

@article{Geiller:2019bti,
    author = {Geiller, Marc and Jai-akson, Puttarak},
    title = {{Extended actions, dynamics of edge modes, and entanglement entropy}},
    journal = {JHEP},
    volume = {09},
    pages = {134},
    year = {2020},
    doi = {10.1007/JHEP09(2020)134},
    eprint = {1912.06025},
    archivePrefix = {arXiv},
    primaryClass = {hep-th}
}

@article{Ghosh:2015iwa,
    author = {Ghosh, Sudip and Soni, Ronak M and Trivedi, Sandip P.},
    title = {{On The Entanglement Entropy For Gauge Theories}},
    journal = {JHEP},
    volume = {09},
    pages = {069},
    year = {2015},
    doi = {10.1007/JHEP09(2015)069},
    eprint = {1501.02593},
    archivePrefix = {arXiv},
    primaryClass = {hep-th},
    reportnumber = {TIFR-TH-15-03}
}

@article{Gibbons:1976ue,
    author = {Gibbons, G. W. and Hawking, S. W.},
    title = {Action integrals and partition functions in quantum gravity},
    journal = {Phys. Rev. D},
    volume = {15},
    pages = {2752--2756},
    year = {1977},
    month = {May},
    publisher = {American Physical Society},
    doi = {10.1103/PhysRevD.15.2752},
    reportnumber = {PRINT-76-0995 (CAMBRIDGE)},
    url = {https://link.aps.org/doi/10.1103/PhysRevD.15.2752},
    issue = {10},
    numpages = {0}
}

@article{Giesel:2024EdgeClocks,
    author = "Giesel, Kristina and Kabel, Viktoria and Wieland, Wolfgang",
    title = "{Linking edge modes and geometrical clocks in linearized gravity}",
    eprint = "2410.17339",
    archivePrefix = "arXiv",
    primaryClass = "gr-qc",
    doi = "10.1103/yrc1-gmql",
    journal = "Phys. Rev. D",
    volume = "112",
    number = "6",
    pages = "064063",
    year = "2025"
}

@article{Gomes:2018dxs,
    author = {Gomes, Henrique and Hopfm\"uller, Florian and Riello, Aldo},
    title = {{A unified geometric framework for boundary charges and dressings: non-Abelian theory and matter}},
    journal = {Nucl. Phys. B},
    volume = {941},
    pages = {249--315},
    year = {2019},
    doi = {10.1016/j.nuclphysb.2019.02.020},
    eprint = {1808.02074},
    archivePrefix = {arXiv},
    primaryClass = {hep-th}
}

@article{Gomes:2019xto,
    author = {Gomes, Henrique and Riello, Aldo},
    title = {{The quasilocal degrees of freedom of Yang-Mills theory}},
    journal = {SciPost Phys.},
    volume = {10},
    number = {6},
    pages = {130},
    year = {2021},
    doi = {10.21468/SciPostPhys.10.6.130},
    eprint = {1910.04222},
    archivePrefix = {arXiv},
    primaryClass = {hep-th}
}

@article{Grewal:2022hlo,
    author = {Grewal, Manvir and Law, Y. T. Albert and Parmentier, Klaas},
    title = {{Black hole horizon edge partition functions}},
    journal = {JHEP},
    volume = {06},
    pages = {025},
    year = {2023},
    doi = {10.1007/JHEP06(2023)025},
    eprint = {2211.16644},
    archivePrefix = {arXiv},
    primaryClass = {hep-th}
}

@article{Harlow:2019yfa,
    author = {Harlow, Daniel and Wu, Jie-Qiang},
    title = {{Covariant phase space with boundaries}},
    journal = {JHEP},
    volume = {10},
    pages = {146},
    year = {2020},
    doi = {10.1007/JHEP10(2020)146},
    eprint = {1906.08616},
    archivePrefix = {arXiv},
    primaryClass = {hep-th}
}

@article{Ivanov:1975zq,
    author = {Ivanov, E. A. and Ogievetsky, V. I.},
    title = {{The Inverse Higgs Phenomenon in Nonlinear Realizations}},
    journal = {Teor. Mat. Fiz.},
    volume = {25},
    pages = {164--177},
    year = {1975},
    doi = {10.1007/BF01028947},
    reportnumber = {JINR-E2-8593}
}

@article{Iyer:1994ys,
    author = {Iyer, Vivek and Wald, Robert M.},
    title = {{Some properties of Noether charge and a proposal for dynamical black hole entropy}},
    journal = {Phys. Rev. D},
    volume = {50},
    pages = {846--864},
    year = {1994},
    doi = {10.1103/PhysRevD.50.846},
    eprint = {gr-qc/9403028},
    archivePrefix = {arXiv}
}

@article{Jiang:2020cqo,
    author = {Jiang, Yikun and Kim, Manki and Wong, Gabriel},
    title = {{Entanglement entropy and edge modes in topological string theory. Part II. The dual gauge theory story}},
    journal = {JHEP},
    volume = {10},
    pages = {202},
    year = {2021},
    doi = {10.1007/JHEP10(2021)202},
    eprint = {2012.13397},
    archivePrefix = {arXiv},
    primaryClass = {hep-th}
}

@article{Joung:2025Extracting,
    author = "Joung, Euihun and Kim, Sejin",
    title = "{Extracting edge modes: Reduction of 3D and 2D gravities}",
    eprint = "2509.09239",
    archivePrefix = "arXiv",
    primaryClass = "hep-th",
    doi = "10.1103/bgjv-m11w",
    journal = "Phys. Rev. D",
    volume = "113",
    number = "2",
    pages = "L021901",
    year = "2026"
}

@article{Kabat:1995eq,
    author = {Kabat, Daniel N.},
    title = {{Black hole entropy and entropy of entanglement}},
    journal = {Nucl. Phys. B},
    volume = {453},
    pages = {281--299},
    year = {1995},
    doi = {10.1016/0550-3213(95)00443-V},
    eprint = {hep-th/9503016},
    archivePrefix = {arXiv},
    reportnumber = {RU-95-06}
}

@article{Kabel:2023BoundaryFrames,
    author = "Kabel, Viktoria and Brukner, {\v{C}}aslav and Wieland, Wolfgang",
    title = "{Quantum reference frames at the boundary of spacetime}",
    eprint = "2302.11629",
    archivePrefix = "arXiv",
    primaryClass = "gr-qc",
    doi = "10.1103/PhysRevD.108.106022",
    journal = "Phys. Rev. D",
    volume = "108",
    number = "10",
    pages = "106022",
    year = "2023"
}

@article{Kapec:2024zdj,
    author = "Kapec, Daniel and Law, Y. T. Albert and Toldo, Chiara",
    title = "{Quasinormal corrections to near-extremal black hole thermodynamics}",
    eprint = "2409.14928",
    archivePrefix = "arXiv",
    primaryClass = "hep-th",
    doi = "10.1007/JHEP06(2025)069",
    journal = "JHEP",
    volume = "06",
    pages = "069",
    year = "2025"
}

@article{Klein:2017npd,
    author = {Klein, Remko and Roest, Diederik and Stefanyszyn, David},
    title = {{Spontaneously Broken Spacetime Symmetries and the Role of Inessential Goldstones}},
    journal = {JHEP},
    volume = {10},
    pages = {051},
    year = {2017},
    doi = {10.1007/JHEP10(2017)051},
    eprint = {1709.03525},
    archivePrefix = {arXiv},
    primaryClass = {hep-th}
}

@article{Klinger:2026tws,
    author = {Klinger, Marc S. and Kudler-Flam, Jonah and Satishchandran, Gautam},
    title = {{Generalized black hole entropy is von Neumann entropy. II. The complete symmetry group and edge modes}},
    eprint = {2601.07910},
    archivePrefix = {arXiv},
    primaryClass = {hep-th},
    doi = {10.1103/tb6y-lsfv},
    journal = {Phys. Rev. D},
    volume = {114},
    number = {6},
    pages = {065010},
    year = {2026}
}

@article{Law:2020cpj,
    author = {Law, Y. T. Albert},
    title = {{A compendium of sphere path integrals}},
    journal = {JHEP},
    volume = {12},
    pages = {213},
    year = {2021},
    doi = {10.1007/JHEP12(2021)213},
    eprint = {2012.06345},
    archivePrefix = {arXiv},
    primaryClass = {hep-th}
}

@article{Law:2022zdq,
    author = "Law, Y. T. Albert and Parmentier, Klaas",
    title = "{Black hole scattering and partition functions}",
    eprint = "2207.07024",
    archivePrefix = "arXiv",
    primaryClass = "hep-th",
    doi = "10.1007/JHEP10(2022)039",
    journal = "JHEP",
    volume = "10",
    pages = "039",
    year = "2022"
}

@article{Law:2023ohq,
    author = {Law, Yuk Ting Albert},
    title = {{Characters, Quasinormal Modes, and Quantum de Sitter Thermodynamics}},
    eprint = {2304.01471},
    archivePrefix = {arXiv},
    primaryClass = {hep-th},
    doi = {10.22323/1.436.0130},
    journal = {PoS},
    volume = {CORFU2022},
    pages = {130},
    year = {2023}
}

@article{Law:2025ktz,
    author = "Law, Y. T. Albert",
    title = "{de Sitter horizon edge partition functions}",
    eprint = "2501.17912",
    archivePrefix = "arXiv",
    primaryClass = "hep-th",
    doi = "10.1103/cm6p-9r4n",
    journal = "Phys. Rev. D",
    volume = "113",
    number = "8",
    pages = "086005",
    year = "2026"
}

@article{Law:2025yec,
    author = "Law, Y. T. Albert and Lochab, Varun",
    title = "{Gravitons on Nariai edges}",
    eprint = "2506.02142",
    archivePrefix = "arXiv",
    primaryClass = "hep-th",
    doi = "10.1103/23vk-l1fl",
    journal = "Phys. Rev. D",
    volume = "113",
    number = "8",
    pages = "086006",
    year = "2026"
}

@article{Law:2026tuk,
    author = {Law, Y. T. Albert and Lochab, Varun},
    title = {{Horizon Edge Partition Functions in {\ensuremath{\Lambda}}{\ensuremath{>}}0 Quantum Gravity}},
    journal = {Phys. Rev. Lett.},
    volume = {136},
    number = {15},
    pages = {151601},
    year = {2026},
    doi = {10.1103/c683-tqw8},
    eprint = {2603.20913},
    archivePrefix = {arXiv},
    primaryClass = {hep-th}
}

@article{Low:2001bw,
    author = {Low, Ian and Manohar, Aneesh V.},
    title = {{Spontaneously broken space-time symmetries and Goldstone's theorem}},
    journal = {Phys. Rev. Lett.},
    volume = {88},
    pages = {101602},
    year = {2002},
    doi = {10.1103/PhysRevLett.88.101602},
    eprint = {hep-th/0110285},
    archivePrefix = {arXiv}
}

@article{Maldacena:2024spf,
    author = {Maldacena, Juan},
    title = {{Real observers solving some imaginary problems}},
    eprint = {2412.14014},
    archivePrefix = {arXiv},
    primaryClass = {hep-th},
    doi = {10.1142/s0217751x2647010x},
    journal = {Int. J. Mod. Phys. A},
    volume = {41},
    number = {26},
    pages = {2647010},
    year = {2026}
}

@article{McArthur:2010zm,
    author = {McArthur, I. N.},
    title = {{Nonlinear realizations of symmetries and unphysical Goldstone bosons}},
    journal = {JHEP},
    volume = {11},
    pages = {140},
    year = {2010},
    doi = {10.1007/JHEP11(2010)140},
    eprint = {1009.3696},
    archivePrefix = {arXiv},
    primaryClass = {hep-th}
}

@article{Mertens:2022ujr,
    author = {Mertens, Thomas G. and Sim\'on, Joan and Wong, Gabriel},
    title = {{A proposal for 3d quantum gravity and its bulk factorization}},
    journal = {JHEP},
    volume = {06},
    pages = {134},
    year = {2023},
    doi = {10.1007/JHEP06(2023)134},
    eprint = {2210.14196},
    archivePrefix = {arXiv},
    primaryClass = {hep-th}
}

@article{Mukherjee:2023ihb,
    author = "Mukherjee, Jyotirmoy",
    title = "{Entanglement entropy and the boundary action of edge modes}",
    eprint = "2310.14690",
    archivePrefix = "arXiv",
    primaryClass = "hep-th",
    doi = "10.1007/JHEP06(2024)113",
    journal = "JHEP",
    volume = "06",
    pages = "113",
    year = "2024"
}

@article{Mukherjee:2025xlt,
    author = "Mukherjee, Jyotirmoy",
    title = "{Quasinormal bulk-edge characters of gravitons in Nariai geometry}",
    eprint = "2506.07556",
    archivePrefix = "arXiv",
    primaryClass = "hep-th",
    doi = "10.1103/bqgb-zzgy",
    journal = "Phys. Rev. D",
    volume = "114",
    number = "2",
    pages = "026034",
    year = "2026"
}

@article{Nicolis:2013lma,
    author = {Nicolis, Alberto and Penco, Riccardo and Rosen, Rachel A.},
    title = {{Relativistic Fluids, Superfluids, Solids and Supersolids from a Coset Construction}},
    journal = {Phys. Rev. D},
    volume = {89},
    number = {4},
    pages = {045002},
    year = {2014},
    doi = {10.1103/PhysRevD.89.045002},
    eprint = {1307.0517},
    archivePrefix = {arXiv},
    primaryClass = {hep-th}
}

@article{Nicolis:2013sga,
    author = {Nicolis, Alberto and Penco, Riccardo and Piazza, Federico and Rosen, Rachel A.},
    title = {{More on gapped Goldstones at finite density: More gapped Goldstones}},
    journal = {JHEP},
    volume = {11},
    pages = {055},
    year = {2013},
    doi = {10.1007/JHEP11(2013)055},
    eprint = {1306.1240},
    archivePrefix = {arXiv},
    primaryClass = {hep-th}
}

@article{Nicolis:2015sra,
    author = {Nicolis, Alberto and Penco, Riccardo and Piazza, Federico and Rattazzi, Riccardo},
    title = {{Zoology of condensed matter: Framids, ordinary stuff, extra-ordinary stuff}},
    journal = {JHEP},
    volume = {06},
    pages = {155},
    year = {2015},
    doi = {10.1007/JHEP06(2015)155},
    eprint = {1501.03845},
    archivePrefix = {arXiv},
    primaryClass = {hep-th}
}

@inproceedings{Ogievetsky1974,
    author = {Ogievetsky, V.I.},
    title = {{Nonlinear Realizations of Internal and Space-time Symmetries}},
    booktitle = {{Proceedings of the X–th Winter School of Theoretical Physics in Karpacz}},
    volume = {1},
    pages = {227},
    year = {1974},
    publisher = {Wroclaw}
}

@article{Parikh:1997ma,
    author = {Parikh, Maulik and Wilczek, Frank},
    title = {{An Action for black hole membranes}},
    journal = {Phys. Rev. D},
    volume = {58},
    pages = {064011},
    year = {1998},
    doi = {10.1103/PhysRevD.58.064011},
    eprint = {gr-qc/9712077},
    archivePrefix = {arXiv},
    reportnumber = {PUPT-1753, IASSNS-HEP-97-117}
}

@article{Polchinski:1988ua,
    author = {Polchinski, Joseph},
    title = {{The Phase of the Sum Over Spheres}},
    journal = {Phys. Lett. B},
    volume = {219},
    pages = {251--257},
    year = {1989},
    doi = {10.1016/0370-2693(89)90387-0},
    reportnumber = {UTTG-31-88}
}

@article{Price:1986yy,
    author = {Price, R. H. and Thorne, K. S.},
    title = {{Membrane Viewpoint on Black Holes: Properties and Evolution of the Stretched Horizon}},
    journal = {Phys. Rev. D},
    volume = {33},
    pages = {915--941},
    year = {1986},
    doi = {10.1103/PhysRevD.33.915}
}

@article{Radicevic:2014kqa,
    author = "Radicevic, Djordje",
    title = "{Notes on Entanglement in Abelian Gauge Theories}",
    eprint = "1404.1391",
    archivePrefix = "arXiv",
    primaryClass = "hep-th",
    reportNumber = "SU-ITP-14-08",
    month = "4",
    year = "2014"
}

@article{Riello:2021lfl,
    author = {Riello, Aldo},
    title = {{Edge modes without edge modes}},
    year = {2021},
    month = {4},
    eprint = {2104.10182},
    archivePrefix = {arXiv},
    primaryClass = {hep-th}
}

@article{Soni:2015yga,
    author = {Soni, Ronak M and Trivedi, Sandip P.},
    title = {{Aspects of Entanglement Entropy for Gauge Theories}},
    journal = {JHEP},
    volume = {01},
    pages = {136},
    year = {2016},
    doi = {10.1007/JHEP01(2016)136},
    eprint = {1510.07455},
    archivePrefix = {arXiv},
    primaryClass = {hep-th},
    reportnumber = {TIFR-TH-15-29}
}

@article{Speranza:2017gxd,
    author = {Speranza, Antony J.},
    title = {{Local phase space and edge modes for diffeomorphism-invariant theories}},
    journal = {JHEP},
    volume = {02},
    pages = {021},
    year = {2018},
    doi = {10.1007/JHEP02(2018)021},
    eprint = {1706.05061},
    archivePrefix = {arXiv},
    primaryClass = {hep-th}
}

@article{Susskind:1993if,
    author = {Susskind, Leonard and Thorlacius, Larus and Uglum, John},
    title = {{The Stretched horizon and black hole complementarity}},
    journal = {Phys. Rev. D},
    volume = {48},
    pages = {3743--3761},
    year = {1993},
    doi = {10.1103/PhysRevD.48.3743},
    eprint = {hep-th/9306069},
    archivePrefix = {arXiv},
    reportnumber = {SU-ITP-93-15}
}

@article{Susskind:2025wyn,
    author = {Susskind, Leonard},
    title = {{Why do we Need Observers? Spontaneous Breaking of Time-Reversal in de Sitter Space}},
    eprint = {2512.13650},
    archivePrefix = {arXiv},
    primaryClass = {hep-th},
    month = {12},
    year = {2025}
}

@article{Susskind:2026fjc,
    author = {Susskind, Leonard},
    title = {{More About the Spontaneous Breaking of Time Reversal in de Sitter Space}},
    eprint = {2601.01666},
    archivePrefix = {arXiv},
    primaryClass = {hep-th},
    doi = {10.22128/jhap.2026.3250.1192},
    journal = {JHAP},
    volume = {6},
    number = {4},
    pages = {1--15},
    year = {2026}
}

@article{Susskind:2026moe,
    author = {Susskind, Leonard},
    title = {{Is Time Reversal in de Sitter Space a Spontaneously Broken Gauge Symmetry?}},
    eprint = {2603.12434},
    archivePrefix = {arXiv},
    primaryClass = {hep-th},
    month = {3},
    year = {2026}
}

@article{Takayanagi:2019tvn,
    author = {Takayanagi, Tadashi and Tamaoka, Kotaro},
    title = {{Gravity Edges Modes and Hayward Term}},
    journal = {JHEP},
    volume = {02},
    pages = {167},
    year = {2020},
    doi = {10.1007/JHEP02(2020)167},
    eprint = {1912.01636},
    archivePrefix = {arXiv},
    primaryClass = {hep-th},
    reportnumber = {YITP-19-116, IPMU19-0173}
}

@article{Volkov:1973vd,
    author = {Volkov, Dmitri V.},
    title = {{Phenomenological Lagrangians}},
    journal = {Fiz. Elem. Chast. Atom. Yadra},
    volume = {4},
    pages = {3--41},
    year = {1973}
}

@article{Wong:2017pdm,
    author = {Wong, Gabriel},
    title = {{A note on entanglement edge modes in Chern Simons theory}},
    journal = {JHEP},
    volume = {08},
    pages = {020},
    year = {2018},
    doi = {10.1007/JHEP08(2018)020},
    eprint = {1706.04666},
    archivePrefix = {arXiv},
    primaryClass = {hep-th}
}

@article{Wong:2022eiu,
    author = {Wong, Gabriel},
    title = {{A note on the bulk interpretation of the quantum extremal surface formula}},
    journal = {JHEP},
    volume = {04},
    pages = {024},
    year = {2024},
    doi = {10.1007/JHEP04(2024)024},
    eprint = {2212.03193},
    archivePrefix = {arXiv},
    primaryClass = {hep-th}
}

@article{Wong:2026ftn,
    author = {Wong, Gabriel},
    title = {{Entanglement and geometric transitions in topological string theory}},
    eprint = {2607.03526},
    archivePrefix = {arXiv},
    primaryClass = {hep-th},
    month = {7},
    year = {2026}
}

@article{Hassan:2011hr,
    author = "Hassan, S. F. and Rosen, Rachel A.",
    title = "{Resolving the Ghost Problem in non-Linear Massive Gravity}",
    eprint = "1106.3344",
    archivePrefix = "arXiv",
    primaryClass = "hep-th",
    doi = "10.1103/PhysRevLett.108.041101",
    journal = "Phys. Rev. Lett.",
    volume = "108",
    pages = "041101",
    year = "2012"
}

@article{deRham:2010kj,
    author = "de Rham, Claudia and Gabadadze, Gregory and Tolley, Andrew J.",
    title = "{Resummation of Massive Gravity}",
    eprint = "1011.1232",
    archivePrefix = "arXiv",
    primaryClass = "hep-th",
    doi = "10.1103/PhysRevLett.106.231101",
    journal = "Phys. Rev. Lett.",
    volume = "106",
    pages = "231101",
    year = "2011"
}

@article{Andrews:2013ora,
    author = "Andrews, Melinda and Goon, Garrett and Hinterbichler, Kurt and Stokes, James and Trodden, Mark",
    title = "{Massive Gravity Coupled to Galileons is Ghost-Free}",
    eprint = "1303.1177",
    archivePrefix = "arXiv",
    primaryClass = "hep-th",
    doi = "10.1103/PhysRevLett.111.061107",
    journal = "Phys. Rev. Lett.",
    volume = "111",
    number = "6",
    pages = "061107",
    year = "2013"
}

@article{deRham:2015ijs,
    author = "de Rham, Claudia and Tolley, Andrew J. and Zhou, Shuang-Yong",
    title = "{Non-compact nonlinear sigma models}",
    eprint = "1512.06838",
    archivePrefix = "arXiv",
    primaryClass = "hep-th",
    doi = "10.1016/j.physletb.2016.07.035",
    journal = "Phys. Lett. B",
    volume = "760",
    pages = "579--583",
    year = "2016"
}

@article{Delacretaz:2014oxa,
    author = "Delacr{\'e}taz, Luca V. and Endlich, Solomon and Monin, Alexander and Penco, Riccardo and Riva, Francesco",
    title = "{(Re-)Inventing the Relativistic Wheel: Gravity, Cosets, and Spinning Objects}",
    eprint = "1405.7384",
    archivePrefix = "arXiv",
    primaryClass = "hep-th",
    doi = "10.1007/JHEP11(2014)008",
    journal = "JHEP",
    volume = "11",
    pages = "008",
    year = "2014"
}

@book{Anand:2020cms,
    author = {Anand, Lallit and Govindjee, Sanjay},
    title = {{Continuum Mechanics of Solids}},
    publisher = {Oxford University Press},
    year = {2020},
    doi = {10.1093/oso/9780198864721.001.0001},
    isbn = {9780198864721},
    note = {Sec.~3.4.1, p.~48}
}

@article{Anninos:2018HolographicLiquids,
    author = {Anninos, Dionysios and Galante, Dami\'an A. and Hofman, Diego M.},
    title = {{De Sitter horizons \& holographic liquids}},
    journal = {JHEP},
    volume = {07},
    pages = {038},
    year = {2019},
    doi = {10.1007/JHEP07(2019)038},
    eprint = {1811.08153},
    archivePrefix = {arXiv},
    primaryClass = {hep-th}
}

@article{Dabholkar:2026afn,
    author = "Dabholkar, Atish and Harris, Eleanor and Moitra, Upamanyu",
    title = "{Edge Modes on Stringy Horizons}",
    eprint = "2601.13131",
    archivePrefix = "arXiv",
    primaryClass = "hep-th",
    doi = "10.1103/cd5m-1knl",
    journal = "Phys. Rev. Lett.",
    volume = "136",
    number = "22",
    pages = "221501",
    year = "2026"
}

@article{Langenscheidt:2024nyw,
    author = "Langenscheidt, Simon and Oriti, Daniele",
    title = "{New edge modes and corner charges for first-order symmetries of 4D gravity}",
    eprint = "2408.01809",
    archivePrefix = "arXiv",
    primaryClass = "gr-qc",
    doi = "10.1088/1361-6382/adbfee",
    journal = "Class. Quant. Grav.",
    volume = "42",
    number = "7",
    pages = "075010",
    year = "2025"
}

@article{He:2024skc,
    author = "He, Temple and Mitra, Prahar and Zurek, Kathryn M.",
    title = "{Diamond of infrared equivalences in abelian gauge theories}",
    eprint = "2405.12303",
    archivePrefix = "arXiv",
    primaryClass = "hep-th",
    reportNumber = "CALT-TH 2024-018",
    doi = "10.1103/PhysRevD.110.105018",
    journal = "Phys. Rev. D",
    volume = "110",
    number = "10",
    pages = "105018",
    year = "2024"
}

@article{Araujo-Regado:2025ejs,
    author = "Araujo-Regado, Goncalo and Hoehn, Philipp A. and Sartini, Francesco",
    title = "{Relational entanglement entropies and quantum reference frames in gauge theories}",
    eprint = "2506.23459",
    archivePrefix = "arXiv",
    primaryClass = "hep-th",
    month = "6",
    year = "2025"
}

@article{Ciambelli:2026pwi,
    author = "Ciambelli, Luca and He, Temple and Klinger, Marc S. and Zurek, Kathryn M.",
    title = "{Mapping the Infrared Phase Space of Gravity to Finite Subregions}",
    eprint = "2606.12515",
    archivePrefix = "arXiv",
    primaryClass = "hep-th",
    reportNumber = "CALT-TH 2026-023",
    month = "6",
    year = "2026"
}

@article{Blommaert:2026ofx,
    author = "Blommaert, Andreas and Tietto, Damiano and Verlinde, Herman",
    title = "{An observer's quantization of 3d de Sitter}",
    eprint = "2606.26241",
    archivePrefix = "arXiv",
    primaryClass = "hep-th",
    month = "6",
    year = "2026"
}

@article{Ciambelli:2024qgi,
    author = "Ciambelli, Luca and Kowalski-Glikman, Jerzy and Varrin, Ludovic",
    title = "{Quantum corner symmetry: Representations and gluing}",
    eprint = "2406.07101",
    archivePrefix = "arXiv",
    primaryClass = "hep-th",
    doi = "10.1016/j.physletb.2025.139544",
    journal = "Phys. Lett. B",
    volume = "866",
    pages = "139544",
    year = "2025"
}

@article{Ciambelli:2025ztm,
    author = "Ciambelli, Luca and Kowalski-Glikman, Jerzy and Varrin, Ludovic",
    title = "{Entanglement entropy of quantum corners}",
    eprint = "2507.16800",
    archivePrefix = "arXiv",
    primaryClass = "hep-th",
    doi = "10.1007/JHEP02(2026)188",
    journal = "JHEP",
    volume = "02",
    pages = "188",
    year = "2026"
}

	\end{document}